\documentclass[a4paper,11pt]{article}
\pdfoutput=1 

\usepackage{jcappub} 
\usepackage{aas_macros}
\usepackage{hyperref}
\usepackage{graphicx}
\usepackage{subcaption}
\usepackage{float}
\usepackage{color}
\usepackage{cancel}
\usepackage{array}
\usepackage{relsize}

\newcommand{\bea}{\begin{eqnarray}}
	\newcommand{\eea}{\end{eqnarray}}

\title{\boldmath E and B modes of the CMB y-type distortions: polarised kinetic Sunyaev-Zeldovich effect from the reionisation and post-reionisation eras}

\author[]{Aritra Kumar Gon,}
\author[]{ Rishi Khatri}

\affiliation[]{ Department of Theoretical Physics, Tata Institute of Fundamental Research, \\
	Mumbai, India}

\emailAdd{aritra.gon@theory.tifr.res.in}
\emailAdd{khatri@theory.tifr.res.in}
\abstract{We study the E and B mode polarisation of the cosmic microwave background (CMB) originating from the transverse peculiar velocity of free electrons, at second order in perturbation theory, during the reionisation and post-reionisation eras. Interestingly, the spectrum of this polarised kinetic Sunyaev-Zeldovich (SZ) effect can be decomposed into a blackbody part and a y-type distortion. The y-distortion part is distinguishable from the primary E and B modes and also the lensing B modes. Furthermore, it is also differentiable from the other y-type signals, such as the thermal SZ effect, which are unpolarised. We show that this signal is sensitive to the reionisation history, in particular to how fast reionisation happens. The E and B modes of y-type distortion provide a way to beat the cosmic variance of primary CMB anisotropies and are an independent probe of the cosmological parameters. The blackbody component of the pkSZ effect would be an important foreground for the primordial tensor modes for tensor to scalar ratio $r \lesssim 3\times10^{-5}$.}

\subheader{TIFR/TH/22-39}
\begin{document}
	\maketitle
	\flushbottom
	\section{Introduction \label{sec_intro}}
	
	In addition to the primary anisotropies created during recombination, several other physical processes at later redshifts can generate secondary anisotropies in the cosmic microwave background (CMB) at linear and higher orders in perturbations. A large number of ongoing and future experiments will measure the polarised CMB anisotropies with progressively higher sensitivity. These experiments will also have a larger number of frequency bands compared to the past experiments. The future experiments (funded and proposed) include ground-based experiments such as the Simons Observatory \cite{ade2019simons}, CMB-S4 \cite{abazajian2016cmb}, and CMB-HD \cite{sehgal2019cmb}  and satellite-based missions such as LiteBIRD \cite{matsumura2014mission}, PIXIE \cite{kogut2011primordial}, PRISM \cite{andre2013prism}, PICO \cite{PICO}, and CMB-Bharat \cite{CMB_Bharat}. This opens up the exciting possibility of having a new window into the Universe using \textit{polarised spectral distortion anisotropies} of the CMB. The secondary anisotropies contain a wealth of information, but detecting them is challenging because of their small amplitude. Even if a CMB experiment has sufficient sensitivity, distinguishing the secondary from the primary anisotropies is difficult if they have the same spectrum. In particular, we are limited by the cosmic variance of the primary CMB anisotropies. The situation becomes more promising if the secondary anisotropies have a different spectrum and thus can evade the cosmic variance limit of the primary and other secondary anisotropies. One such physical process is the second order polarisation of the CMB due to the kinetic Sunyaev-Zeldovich (pkSZ) effect which is the focus of this paper. This pkSZ effect was first predicted in 1980 by Rashid Sunyaev and Yakov Zeldovich \cite{SZ_80}. The pkSZ effect from reionisation as well as from galaxy clusters has been studied previously \cite{Sazonov1999,Hu_2000, BAUMANN2003, roebber2014polarization,Pierpaoli2016}. Our work differs from the previous studies in several ways. Instead of a flat sky approximation, we derive the full sky exact expressions of the E and B mode power spectrum of the pkSZ effect. We also study the effect of different reionisation histories on the power spectrum, instead of assuming instantaneous reionisation. We show that the pkSZ effect is sensitive to the reionisation history and future observations can in principle extract information about various cosmological parameters, beating the cosmic variance limit of primary anisotropies. A similar analysis was previously done by Renaux-Petel et al. \cite{renaux2014spectral}. We compare our results with theirs in section \ref{sec_result}.
	
	During the era of reionisation (z $\sim 6-20$) \cite{adam2016planck}, the free electrons that are produced have some peculiar velocity with respect to the CMB rest frame. As a result, in the electron rest frame, the CMB is no longer isotropic \cite{kamionkowski2003aspects, peebles_fireball}. In addition to the dipole, multipoles of all higher orders are present in the intensity of incoming radiation in the electron rest frame. This occurs due to the non-linear nature of the relativistic Doppler boost, as well as the non-linear relation between the temperature and the intensity in the Planck spectrum. In particular, a quadrupolar anisotropy gets generated. Thomson scattering of this quadrupolar anisotropic radiation by the electrons produces linear polarisation \cite{dodelson2003modern, durrer2020cosmic, KOSOWSKY199649,chandrasekhar2013radiative}. The polarisation strength is proportional to the square of the transverse velocity of the electrons. More importantly, the spectral signature of this polarised signal is different from that of primary CMB polarisation as well as the lensing B modes. It can be shown that the intensity quadrupole consists of a blackbody spectrum along with a y-type distortion \cite{Sazonov1999,kamionkowski2003aspects,Sunyaev_2013, Chluba_superposition_BB}. The y-type distortion (SZ spectrum) provides a unique signature to this polarisation signal which makes it possible to detect using component separation techniques, provided the required sensitivity is achieved in future. The y-distortion part is also free from the cosmic variance of the primary polarisation as they do not have the same spectrum. Since the signal is generated at higher order in perturbation theory \cite{fidler2014intrinsic}, we expect to get both E and B modes even when the velocity fields are sourced by purely scalar perturbations. We perform a full sky numerical calculation of the y-type angular power spectrum of both the E and B modes. We assume a homogeneous electron density during reionisation which evolves with redshift. We include both symmetric and asymmetric reionisation \cite{lewis2008cosmological, adam2016planck}. For completeness, we also include the second reionisation of helium \cite{heinrich2017complete}. We show that the pkSZ effect is sensitive to the central redshift as well as the duration of reionisation. If detected, this signal can be instrumental in distinguishing different reionisation histories. It can also act as an independent probe to measure large-scale velocity fields and thus constrain other cosmological parameters. The blackbody part of the signal will also act as a foreground for the primordial B modes. A precise measurement of the intrinsic B mode polarisation has important implications for understanding the physics of the early universe \cite{baumann2009probing, krauss2010primordial}. In order to correctly measure the primordial signal, an accurate prediction of these foreground signals is necessary.
	
\begin{figure}
		\hspace{-0.9cm}
		\begin{subfigure}{1.0\textwidth}
		\centering
		\includegraphics[width=1.07\linewidth]{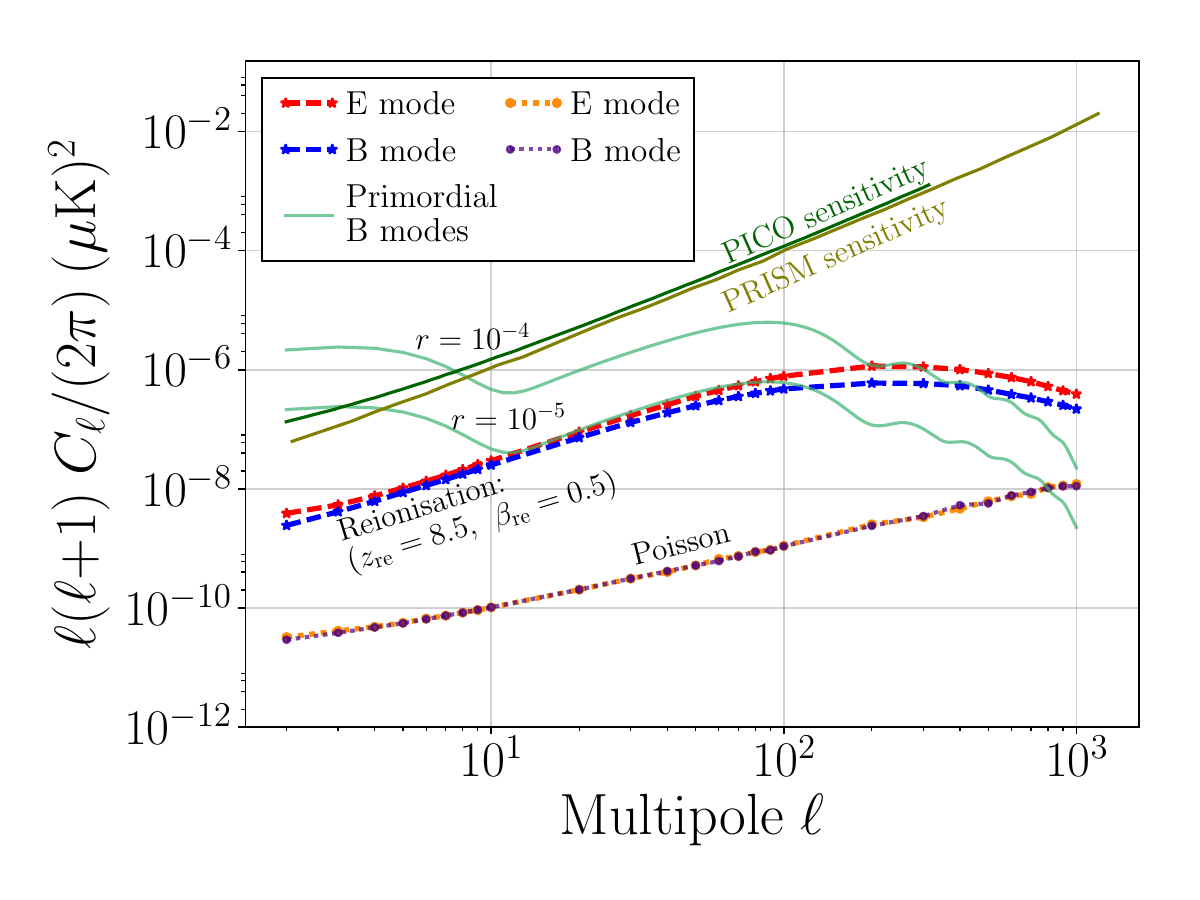}
	\end{subfigure}
		\caption{The dashed curves are the y-type E and B mode polarisation angular power spectrum in temperature units using the conversion $\Delta \mathrm{T}=$2(y-amplitude)$\mathrm{T_{CMB}}$, corresponding to the Rayleigh-Jeans part of the CMB spectrum. The solid curves are the primordial B modes at $r= 10^{-4}$ and $r=10^{-5}$. The dotted lines are the E and B modes from the galaxy clusters contributed by Poisson or shot noise. We also show the sensitivity of the proposed space mission PRISM \cite{andre2013prism} and PICO \cite{PICO} for reference. The blackbody component of the pkSZ E and B mode power spectrum is
		a factor of 4 larger. The B-mode power in the blackbody component is equal to the primordial B-mode power spectrum for $ r= 3.0\times10^{-5}$ at $\ell=100$.}
\label{fig:summary plot}
	\end{figure}

	We show the E and B mode power spectrum, for our fiducial symmetric reionisation history chosen to be consistent with current observations, in figure \ref{fig:summary plot}. Also shown is the Poisson noise contribution due to the galaxy clusters. For comparison, the primordial B modes for tensor to scalar ratios $r$ of $10^{-4}$ and $10^{-5}$ are given. We have ignored the spatial variation in the electron density field. As we see, the low redshift contribution from the Poisson noise from galaxy clusters is $\sim$ 2 orders of magnitudes smaller than the reionisation signal. The PRISM sensitivity curve is plotted to provide an idea about the detectability of this signal in future experiments. More details are given in section \ref{sec_result}.
	We will assume a flat $\mathrm{\Lambda CDM}$ universe with baryon and matter density parameters $\mathrm{\Omega_b = 0.0490}$ and $\mathrm{\Omega_m = 0.3111}$, Hubble constant, $\mathrm{H_0} =$ 100h $\mathrm{kms^{-1} Mpc^{-1}}$ with $\mathrm{h = 0.6766}$, spectral index of primordial curvature perturbations $\mathrm {ns = 0.9665}$, its amplitude $\mathrm {log(As)} =-8.678 $ and helium mass fraction $\mathrm{X_{He} = 0.24}$ \cite{aghanim2020planck}. We used publicly accessible codes CAMB \cite{lewis2000efficient} and Colossus \cite{diemer2018colossus} for our numerical analysis and Vegas \cite{lepage1978new} for multidimensional adaptive Monte-Carlo integration. We will be using units with the speed of light in vacuum $c=1$.
	\section{Polarised kinetic Sunyaev-Zeldovich effect \label{sec_polfield}}
	A general photon distribution can be characterised by a set of Stokes parameters, $\mathrm{I,\;Q}$, $\mathrm{U}$, and $\mathrm{V}$, where I is the intensity, Q and U measure the linear polarisation, and V is the measure of the circular polarisation.  Since the Thomson scattering does not generate circular polarisation \cite{rybicki1991radiative,dodelson2003modern,chandrasekhar2013radiative}, we can define a triad $\mathcal{T}$ to describe the incoming and outgoing radiation in a Thomson scattering process,
	\bea
	\mathcal{T}=\left\{\frac{\delta \mathrm{I}}{\mathrm{I}}, \frac{ \mathrm{Q}}{\mathrm{I}},\frac {\mathrm{U}}{\mathrm{I}}\right\}=\{\mathcal{I},\mathcal{Q},\mathcal{U}\},
	\eea
	where $\mathcal{I}$ is the average background intensity and $\delta \mathrm{I}$ is the difference in intensity with respect to the background. In the case of the CMB, we are interested in the polarisation anisotropies and must work with the polarisation field. The Stokes parameters now are dependent on position, time, and the momentum of the photons. We will use the combination $\left(\mathcal{Q}\pm i\mathcal{U}\right)\left(\mathbf{ r},\mathbf{p},\eta\right)\equiv P_{\pm}\left(\mathbf{ r},\mathbf{p},\eta\right)$ to describe the 3D polarisation field, where $\mathbf{ r}$ is the position vector, $\mathbf{ p}$ is the momentum vector, and $\eta$ is the conformal time. The momentum vector is given by $\mathbf{ p}=\mathrm{p}\mathbf{\hat{n}}$, where p is the magnitude and $\mathbf{\hat{n}}$ is the direction of propagation. The full photon distribution can be decomposed into a spectral shape and an amplitude. For the CMB, the spectral dependence on the momentum p is known separately and can be factored out of the Boltzmann equation \cite{dodelson2003modern, Pitrou_y_sky, Chluba_2by2}. Hence, we only need to evolve the amplitude part of a particular spectral shape. Therefore, the polarisation field becomes a function of $\mathbf{ r}, \mathbf{\hat{n}}$ and $\eta$. The time evolution of this polarisation field is given by the Boltzmann equation \cite{Seljak_1996, durrer2020cosmic} 
	\bea
	\frac{dP_{\pm}(\mathbf{ r},\mathbf{\hat{n}},\eta)}{d\eta}=C[P_{\pm}],
	\eea
	where $\hat{n}$ is the direction of incoming photons and $\eta$ is the conformal time and $C[P_\pm]$ is the Thomson collision term which can be written as  \cite{Hu_White_97,2013_Tram}
	\bea
	C[P_{\pm}]=\tau'P_{\pm}(\mathbf{ r},\mathbf{\hat{n}},\eta)-\tau'P^{\pm}_{\mathrm{sc}}\left(\mathbf{ r},\mathbf{\hat{n}},\eta\right),
	\eea
	where the first term on the right hand side accounts for the photons that are scattered out of the line of sight, while the second term is the source term.  The differential optical depth or the scattering rate is given by
	 \bea
	 \tau'=\frac{d\tau}{d\eta}=-n_\mathrm{e}a\sigma_{\mathrm{T}},
	 \eea
	 where $n_\mathrm{e}$ is the electron number density, $a$ is the expansion scale factor, and $\sigma_{\mathrm{T}}$  is the Thomson scattering cross section. The source term $P_{\mathrm{sc}}\left(\mathbf{ r},\mathbf{\hat{n}},\eta\right)$ is given by \cite{chandrasekhar2013radiative, durrer2020cosmic, piattella2018lecture}
	\bea
	\label{p_in}
	P^{\pm}_{\mathrm{sc}}\left(\mathbf{ r},\mathbf{\hat{n}},\eta\right)=-\frac{\sqrt{6}}{10}\sum_{\lambda =-2}^{2}\,_{\pm2}Y_{2 \lambda}\left(\mathbf{  \hat{n}}\right)\int d^{2}\mathbf{ \hat{n}}\;Y_{2 \lambda}^{*}\left(\mathbf{  \hat{n}'}\right)\mathcal{I}_{\mathrm{sc}}\left(\mathbf{r},\mathbf{  \hat{n}'},\eta\right),
	\eea
	where $\mathbf{ \hat{n}'}$ is the incoming photon direction in the rest frame of the electrons and $\mathcal{I}_{\mathrm{sc}}\left(\mathbf{r},\mathbf{  \hat{n}'},\eta\right)$ is the corresponding intensity of the incoming unpolarised blackbody CMB radiation. The  spin-2 and spin-0 spherical harmonic functions are $\,_{\pm2}Y_{2 \lambda}$ and $Y_{2 \lambda}$ respectively. The orthogonality condition of spherical harmonics ensures that the integral in the above equation is non-zero only if $\mathcal{I}_{\mathrm{sc}}\left(\mathbf{r},\mathbf{  \hat{n}'},\eta\right)$ has a quadrupolar anisotropy. When solving the Boltzmann equation, we integrate along our line of sight direction. This implies that the position vector $\mathbf{r}$ along the line of sight is a function of $\mathbf{  \hat{n}}$ and $\eta$. We consider  $\eta_0$ to be the conformal time today and $\eta_{i}$ to be the conformal time at some early epoch before reionisation. Also, at $\eta_0$, $\tau(\eta_0)=0$, by the definition of $\tau$,
	\bea
	\label{tau_chi}
	\tau(\eta)=\int_{\eta}^{\eta_0}n_\mathrm{e}(\eta)\sigma_{\mathrm{T}}a\,d\eta.
	\eea
	Therefore, the polarisation field today can be written as \cite{Khatri_crinkles}
	\bea
	\label{b_soln}
	P_{\pm}\left(\mathbf{  \hat{n}};\eta_0\right)=e^{-\tau(\eta_{i})}P_{\pm}\left(\mathbf{  \hat{n}};\eta_{i}\right)-\int_{\eta_{i}}^{\eta_0}e^{-\tau(\eta)}\tau'P^{\pm}_{\mathrm{sc}}\left(\mathbf{ r},\mathbf{  \hat{n}},\eta\right)d\eta,
	\eea
	where the integral is performed along the line of sight, $\mathbf{ r}=\mathbf{ r}(\eta,\mathbf{\hat{n}})$. We are interested in the scattering of initially unpolarised CMB radiation. Thus we have, $P\left(\mathbf{  \hat{n}};\eta_{i}\right)=0$.  Therefore, using eq.(\ref{p_in}) in eq.(\ref{b_soln}) we get
	\bea
	P_{\pm}\left(\mathbf{  \hat{n}};\eta_0\right)=\int_{\eta_{i}}^{\eta_0}d\eta \;\frac{\sqrt{6}}{10}\tau'e^{-\tau(\eta)}\sum_{\lambda =-2}^{2}\,_{\pm2}Y_{2 \lambda}\left(\mathbf{ \hat{n}}\right)\int d^{2}\mathbf{ \hat{n}'}\;Y_{2 \lambda}^{*}\left(\mathbf{ \hat{n}'}\right)\mathcal{I}_{\mathrm{sc}}\left(\mathbf{r},\mathbf{  \hat{n}'},\eta\right).
	\eea
	We can do a change of variables from conformal time $(\eta)$ to comoving distance $(\chi=\eta_0-\eta)$, to get the polarisation field at $\chi=0$,
	\begin{align}
	\label{pol_sem_final}
	P_{\pm}\left(\hat{\mathbf{ n}}\right)=-\frac{\sqrt{6}\sigma_{\mathrm{T}}}{10}\sum_{\lambda =-2}^{2}\int_{0}^{\chi_{i}}d\chi\;e^{-\tau(\chi)}\,n_\mathrm{e}(\chi)a(\chi)\,_{\pm2}Y_{2 \lambda}\left(\mathbf{ \hat{n}}\right)\int d^{2}\mathbf{ \hat{n}'}\;Y_{2 \lambda}^{*}\left(\mathbf{ \hat{n}'}\right)\mathcal{I}_{\mathrm{sc}}\left(\mathbf{r},\mathbf{  \hat{n}'},\chi\right),
	\end{align}
	where $(\chi_{i}=\eta_0-\eta_{i})$. Since we are considering homogeneous reionisation, the electron density field has no spatial variation. It is a function of redshift (comoving distance) only. The explicit form of reionisation history will be described in section \ref{sec_EB_pow}. To complete the discussion, we need to know the spectrum of the polarised radiation after scattering.
	\subsection{Spectral signature \label{sec_spectral}}
	The origin of the quadrupole and the spectral distortion of incoming CMB intensity in the electron frame,  $\mathcal{I}_{\mathrm{sc}}\left(\mathbf{r},\mathbf{  \hat{n}'},\chi\right)$, can be understood by looking at the Doppler boost when we shift to electron rest frame from the CMB rest frame. In the electron frame, the CMB photons coming from different directions follow a blackbody spectrum with temperature in direction $\mathbf{\hat{n}}$  given by \cite{rybicki1991radiative, Sunyaev_2013} 
	\begin{align}
		\label{ksz_temp}
		T\left(\mathbf{ r},\mathbf{ \hat{n}'},\chi\right)=\frac{T_0(\chi)}{\gamma\left(1+\mathbf{  v}(\mathbf{r},\chi)\cdot\mathbf{  \hat{n}'}\right)}&=T_0(\chi)\left[1\underbrace{-\mathbf{  v}\cdot\mathbf{  \hat{n}'}+\frac{1}{2}v^2+\left(\mathbf{  v}\cdot\mathbf{  \hat{n}'}\right)^2+\mathcal{O}\left(v^3\right)+\cdot\cdot\cdot}_{\theta(\mathbf{ \mathbf{r}, \hat{n}'},\chi)}\right]\nonumber\\
		&\equiv T_0(\chi)\left[1+\theta(\mathbf{ \mathbf{r}, \hat{n}'},\chi)\right],
	\end{align}
	where $T_0(\chi)$ is the average temperature of the CMB, the velocity field of the electrons with respect to the CMB rest frame is given by $\mathbf{  v}(\mathbf{r},\chi)$, and $\gamma=1/\sqrt{\left(1-v^2\right)}$ is the Lorentz factor associated with the transformation from the CMB rest frame to the electron rest frame. Due to the non-linearity of the relativistic Doppler shift, if we expand the temperature in a Taylor series, we see that the multipole moments of all orders are present. The spectrum still remains a blackbody in each direction. The blackbody spectrum however is also a non-linear function of the temperature. Further expanding the intensity in a Taylor series and subtracting the average background blackbody spectrum, we see that the spectrum at second order is no longer a pure blackbody but has a y-type component too. This is essentially the y-type distortion produced by the mixing of blackbodies in the Thomson scattering \cite{Illarionov_mixing_BB_1972, Chluba_superposition_BB, khatri_mixing_BB}. 
	Therefore, the difference in intensity $\delta I_{\nu}=I_{\nu}-\bar{I}_{\nu}$ or the difference in occupation number $\delta n_{\nu}=n_{\nu}-\bar{n}_{\nu}$ with respect to the average background as seen by the electron is given by
	\begin{align}
	\delta n_{\nu}=\frac{1}{2h\nu^{3}}\delta I_{\nu}=\left(\theta+\theta ^2\right)\left(T\frac{\partial n_{pl}}{\partial T}\right)\bigg|_{T_0}+\frac{\theta^2}{2}\left(T^4\frac{\partial}{\partial T}\left(\frac{1}{T^2}\frac{\partial n_{pl}}{\partial T}\right)\right)\bigg|_{T_0}+\mathcal{O}(\theta^3)\;\cdots,
	\end{align}
	where $I_{\nu}$ and $n_{\nu}$ are the intensity and occupation number in the electron rest frame and $\bar{I}_{\nu}$ and $\bar{n}_{\nu}$ are the intensity and occupation number of the average blackbody spectrum with temperature $T_0$. The resultant fractional relative intensity is given by
	\begin{align}
	\label{temp_quadpole}
	\frac{\delta I}{I}\equiv \mathcal{I}_{\mathrm{sc}}=\frac{\delta n_{\nu}}{n_{\nu}}=\left(\theta+\theta ^2\right)g(x)+\frac{\theta^2}{2}y(x)+\mathcal{O}(\theta^3)\:\:\cdots,
	\end{align}
	where $g(x)=\frac{xe^x}{(e^x-1)}$ is the differential blackbody spectrum, $y(x)=\frac{xe^x}{(e^x-1)}\left(x\frac{e^x+1}{e^x-1}-4\right)$ is the y-type distortion spectrum, and $x=\left(\frac{h\nu}{k_BT_o}\right)$ is the dimensionless frequency. We note that $g(x)$ is also the spectrum of the primordial CMB anisotropies for all the CMB experiments which make a differential measurement of the CMB.
	We are interested in the quadrupolar component which will contribute to the polarisation. Collecting the terms from eq.(\ref{temp_quadpole}) which will contribute to the quadrupolar moment we get
	\bea
	\label{k_quadrupole}
	\mathcal{I}_{\mathrm{sc}}\big|_{(\mathrm{quadrupolar})}=2\left(\mathbf{ v}\cdot\mathbf{ \hat{n}'}\right)^2g(x)+\frac{1}{2}\left(\mathbf{ v}\cdot\mathbf{ \hat{n}'}\right)^2y(x).
	\eea
	It is important to note that the quadrupole consist of a blackbody spectrum with an amplitude,  $2\left(\mathbf{ v}\cdot\mathbf{ \hat{n}'}\right)^2$ along with a y-type distortion with an amplitude (y-amplitude),  $\frac{1}{2}\left(\mathbf{ v}\cdot\mathbf{ \hat{n}'}\right)^2$ \cite{zeldovich1969interaction}. It is because of this y-type (SZ type) distortion we will be able to distinguish the pkSZ effect from other signals. We should emphasize that there are two different sources of this quadrupole. One contribution is due to the non-linear nature of the relativistic Doppler shift itself which creates a temperature quadrupole. This only contributes to the blackbody part of the scattered polarised spectrum. The second quadrupole arises due to the non-linear relation between the intensity and the temperature. This contributes to both the blackbody part and the SZ spectrum.  Since eq.(\ref{k_quadrupole}) fixes the spectrum of the scattered radiation, we only need to calculate the amplitude. Therefore, we can replace $\mathcal{I}_{\mathrm{sc}}\left(\mathbf{r},\mathbf{ \hat{n}'},\chi\right)$ by $\left(\mathbf{ v}(\mathbf{r},\chi)\cdot\mathbf{ \hat{n}'}\right)^2$ in eq.(\ref{pol_sem_final}) with an understanding that the spectrum is given by $2g(x)+\frac{1}{2}y(x)$. So, the expression for polarisation field becomes,
	\begin{align}
	\label{pol_final}
	P_{\pm}\left(\hat{\mathbf{ n}}\right)=-\frac{\sqrt{6}\sigma_{\mathrm{T}}}{10}\sum_{\lambda =-2}^{2}\int_{0}^{\chi_{i}}d\chi\;e^{-\tau(\chi)}\,n_\mathrm{e}(\chi)\,a(\chi)\,_{\pm2}Y_{2 \lambda}\left(\mathbf{ \hat{n}}\right)\int d^{2}\mathbf{ \hat{n}'}\;Y_{2 \lambda}^{*}\left(\mathbf{ \hat{n}'}\right)\left(\mathbf{ v}(\mathbf{r},\chi)\cdot\mathbf{ \hat{n}'}\right)^2.
	\end{align}
	The expression $\sum_{\lambda =-2}^{2}\,_{2}Y_{2\lambda}\left(\mathbf{ \hat{n}}\right)\int d^{2}\mathbf{ \hat{n}'}\;Y_{2 \lambda}^{*}\left(\mathbf{ v}\cdot\mathbf{ \hat{n}'}\right)^2$ reduces to the square of transverse velocity $v_t$  along with some numerical factors and a phase (as shown in Appendix \ref{App:Quad_dep}). Thus only the transverse to the line of sight component of the velocity contributes to the polarisation signal and is proportional to the square of the transverse velocity field as expected \cite{SZ_80}. Having checked this explicitly we proceed with our calculations with the total electron velocity field. We will mostly be interested in the y-type part of the signal and will present all our results as E and B mode power spectrum of the y-type distortion. In order to compare with CMB polarisation signals, we will present our results in temperature units using the conversion $\Delta T=$2(y-amplitude)$T_{\mathrm{CMB}}$ for the y-distortion valid in the Rayleigh-Jeans (RJ) part of the spectrum. The blackbody part of the spectrum is twice this value, i.e. $\Delta T_{\mathrm{BB}}=2 \Delta T$. We can now proceed to extract the harmonic coefficients of E and B modes.
	\section{E and B mode harmonic coefficients \label{sec_EB_harm}}
	In the helicity basis, the  pair  $\left(\mathcal{Q}\pm i\mathcal{U}\right)\left(\mathbf{\hat{n}}\right)\equiv	P_{\pm}\left(\mathbf{\hat{n}}\right)$ transforms as a spin-2 field under rotation about $\mathbf{\hat{n}}$. Thus, on a 2-sphere, we can decompose  $	P_{+}\left(\mathbf{\hat{n}}\right)$  as
	\bea
	\label{defn_Q+iU}
		P_{+}\left(\mathbf{\hat{n}}\right)=\sum_{\ell,m}a_{\ell m}\,\,_{2}Y_{\ell m}\left(\mathbf{\hat{n}}\right).
	\eea
	Therefore
	\bea
	\label{harm_coeff}
	a_{\ell m}=\int	P_{+}\left(\mathbf{\hat{n}}\right)\,_{2}Y^*_{\ell m}\left(\mathbf{\hat{n}}\right)d^2{\hat{\mathbf{n}}},
	\eea
	where $\,_{\pm2}Y_{\ell m}$ are the spin-2 spherical harmonic functions. The sum over $\ell$ starts from $\ell=2$ as the spin weighted spherical harmonics,  $\,_{\pm s}Y_{\ell m}$ vanishes for $|s|>l$. We can obtain $	P_{-}\left(\mathbf{\hat{n}}\right)$ by complex conjugation of $	P_{+}\left(\mathbf{\hat{n}}\right)$.
	\begin{align}
		P_{-}\left(\mathbf{\hat{n}}\right)&=\sum_{\ell,m}a^{*}_{\ell m}\,\,_2Y^{*}_{\ell m}\left(\mathbf{\hat{n}}\right)
		=\sum_{\ell,m}a^{*}_{\ell -m}\,(-1)^{m}\,_{-2}Y_{\ell m}\left(\mathbf{\hat{n}}\right).
	\end{align}
	We can now define the E and B mode coefficients as
	\bea
	\label{e_b_coeff}
	e_{\ell m}=\frac{1}{2}\left(a_{\ell m}+(-1)^{m}a^{*}_{\ell -m}\right),\hspace{0.5in}b_{\ell m}=-\frac{i}{2}\left(a_{\ell m}-(-1)^{m}a^{*}_{\ell -m}\right).
	\eea
	Therefore we can write,
	\bea
     P_{\pm}\left(\mathbf{\hat{n}}\right)=\sum_{\ell,m}\left(e_{\ell m}\pm i\,b_{\ell m}\right)\,\,_{\pm2}Y_{\ell m}\left(\mathbf{\hat{n}}\right).
	\eea 
	Since the Stokes parameters $\mathcal{Q}$ and $\mathcal{U}$  are not coordinate invariant, we define scalar fields which, like temperature perturbation, will be coordinate invariant quantities. This can be achieved using spin raising ($\cancel{\partial}$) and lowering operators ($\cancel{\partial}^{*}$) which creates ordinary spherical harmonics from spin weighted harmonic functions \cite{durrer2020cosmic}. \\
	\bea
	\cancel{\partial}^{2}(\,_{-2}Y_{\ell m})=\sqrt{\frac{(\ell +2)!}{(\ell -2)!}}\;Y_{\ell m}\;,
	\hspace{1in}
	\left(\cancel{\partial}^{*}\right)^{2}(\,_{2}Y_{\ell m})=\sqrt{\frac{(\ell +2)!}{(\ell -2)!}}\;Y_{\ell m}.
	\eea
	The $\mathcal{E}$ and $\mathcal{B}$ fields  are related to $\mathcal{Q}$ and $\mathcal{U}$ as
	\begin{align}
		\mathcal{E}(\mathbf{\hat{n}})=\frac{1}{2}\left[\left(\cancel{\partial}^{*}\right)^{2}P_{+}\left(\mathbf{\hat{n}}\right)+\left(\cancel{\partial}\right)^{2}P_{-}\left(\mathbf{\hat{n}}\right)\right]
		=\sum_{\ell,m}e_{\ell m}\sqrt{\frac{(\ell +2)!}{(\ell -2)!}}\,Y_{\ell m}\left(\mathbf{\hat{n}}\right)	\label{Emode_real}
	\end{align}
	and
	\begin{align}
		\mathcal{B}(\mathbf{\hat{n}})=-\frac{i}{2}\left[\left(\cancel{\partial}^{*}\right)^{2}P_{+}\left(\mathbf{\hat{n}}\right)-\left(\cancel{\partial}\right)^{2}P_{-}\left(\mathbf{\hat{n}}\right)\right]
		=\sum_{\ell,m}b_{\ell m}\sqrt{\frac{(\ell +2)!}{(\ell -2)!}}\,Y_{\ell m}\left(\mathbf{\hat{n}}\right)\label{Bmode_real}.
	\end{align} 
	To find the E and B modes, we first convert eq.(\ref{pol_final}) to Fourier space. Since the velocity fields are sourced by scalar modes, we can write $\mathbf{v}\left(\mathbf{r},\chi\right)=\nabla u\left(\mathbf{r},\chi\right)$, where $u$ is the velocity potential. Therefore in Fourier space we have, $\mathbf{\tilde{v}}(\mathbf{k},\chi)=-i\,\mathbf{\hat{k}}\,\tilde{u}(\mathbf{k},\chi)$, where
	\begin{align}
	\label{window_0}
	\mathbf{v}\left(\mathbf{r},\chi\right)=\int \frac{d^3\mathbf{k}}{(2\pi)^3}\mathbf{\tilde{v}}(\mathbf{k},\chi)\;e^{i\mathbf{k}\cdot \mathbf{r}}.
	\end{align}
	From here onward, we will suppress the $\chi$ dependence in $\mathbf{\tilde{v}}$ and $\tilde{u}$. The scalar product between the electron velocity and the incoming photon direction transforms as
	\begin{align}
	\label{vdotn_sq_f_space}
	\left(\mathbf{\tilde{v}}(\mathbf{k_{1}})\cdot\mathbf{\hat{n}}'\right)\left(\mathbf{\tilde{v}}(\mathbf{k_{2}})\cdot\mathbf{\hat{n}}'\right)=- \tilde{u}(\mathbf{k_{1}})\tilde{u}(\mathbf{k_{2}})\left(\mathbf{\hat{k}_{1}}\cdot\mathbf{\hat{n}}'\right)\left(\mathbf{\hat{k}_{2}}\cdot\mathbf{\hat{n}}'\right).
	\end{align}
	We can now perform the integrals over $d^{2}\mathbf{ \hat{n}'}$, using relations between spherical harmonics and scalar product of two vectors to get
	\begin{align}
		\label{n'_integral}
		\int d^{2}\mathbf{\hat{n}'}\; Y_{2\lambda}^{*}\left(\mathbf{\hat{n}'}\right)\left(\mathbf{\hat{k}_1}\cdot\mathbf{\hat{n}'}\right)\left(\mathbf{\hat{k}_2}\cdot\mathbf{\hat{n}'}\right)=(-1)^\lambda \left(\frac{4\pi}{3}\right)^2\sqrt{\frac{3}{2\pi}}\; \sum_{p_1,p_2}\left(\begin{array}{ccc}
			1& 1 & 2\\ 
			p_1& p_2 & -\lambda
		\end{array}\right)Y_{1p_1}^{*}(\mathbf{{\hat{k}}_1})Y_{1p_2}^{*}(\mathbf{{\hat{k}_2}}),
	\end{align}
	where $\left(\begin{array}{ccc}
		l_2& l_3 & l_1\\ 
		m_2& m_3 & -m_1
	\end{array}\right)$ is the Wigner 3j symbol \cite{varshalovich1988quantum}. Using eq.(\ref{vdotn_sq_f_space}) and eq.(\ref{n'_integral}) in eq.(\ref{pol_final}) we obtain the following expression for $P_{+}\left(\hat{\mathbf{ n}}\right)$:
	\begin{align}	
		\label{pol_ksp}
		P_{+}\left(\hat{\mathbf{ n}}\right)=\left(\frac{4\pi}{3}\right)^2&\sqrt{\frac{3}{2\pi}}\frac{\sqrt{6}\sigma_{\mathrm{T}}}{10}\sum_{\lambda =-2}^{2}(-1)^\lambda \int_{0}^{\chi_{i}}d\chi\;e^{-\tau(\chi)}\,n_\mathrm{e}(\chi)\,a(\chi)\,_{2}Y_{2 \lambda}\left(\mathbf{ \hat{n}}\right)\nonumber\\
		&\int \int \frac{d^3\mathbf{k_1}d^3\mathbf{k_2}}{(2\pi)^6} e^{i\left(\mathbf{k_1}+\mathbf{k_2}\right)\cdot \mathbf{r}}\; \tilde{u}(\mathbf{k_{1}})\tilde{u}(\mathbf{k_{2}})\sum_{p_1,p_2}
		\left(\begin{array}{ccc}
			1& 1 & 2\\ 
			p_1& p_2 & -\lambda
		\end{array}\right)Y_{1p_1}^{*}(\mathbf{{\hat{k}}_1})Y_{1p_2}^{*}(\mathbf{{\hat{k}_2}}).
	\end{align}
	Since $P_{-}\left(\hat{\mathbf{ n}}\right)$ is related to $P_{+}\left(\hat{\mathbf{ n}}\right)$ through a complex conjugation, we only need to consider $P_{+}\left(\hat{\mathbf{ n}}\right)$ for our calculation. Defining  $\mathbf{k}=\mathbf{k_1}+\mathbf{k_2}$, we expand the exponential in eq.(\ref{pol_ksp}) into spherical harmonics and spherical Bessel functions $j_\ell(x)$ using the identity,
	\bea
	\label{exp_ylm}
	\exp\left(i\mathbf{k}\cdot\mathbf{r}\right)=4\pi\sum_{L,M}i^L\,Y_{LM}^{*}(\hat{\mathbf{k}})Y_{LM}(\mathbf{\hat{n}})\,j_{L}(k\chi).
	\eea
Substituting in eq.(\ref{harm_coeff}) and using \cite{varshalovich1988quantum, durrer2020cosmic}
	\begin{align}
		\label{2ylm}
		\int Y_{LM}(\mathbf{\hat{n}})\,_{2}Y_{2 \lambda}\left(\mathbf{\hat{n}}\right)\,_{2}&Y^*_{\ell m}(\mathbf{\hat{n}})\,d^2{\hat{\mathbf{n}}}=\nonumber\\ &\sqrt{\frac{{(2L+1)}(2.2+1)(2\ell +1)}{4\pi}}(-1)^{(m)}
		\left(\begin{array}{ccc}
			L& 2 & \ell\\ 
			0& -2 & 2
		\end{array}\right)
	\left(\begin{array}{ccc}
		L& 2& \ell\\ 
		M& \lambda & -m
	\end{array}\right),
	\end{align}
	we get
	\begin{align}
		\label{harm_coeff_start}
		a_{\ell m}=&4\pi\left(\frac{4\pi}{3}\right)^2\sqrt{\frac{3}{2\pi}}\frac{\sqrt{6}\sigma_{\mathrm{T}}}{10}\sum_{\lambda =-2}^{2}(-1)^\lambda \int_{0}^{\chi_{i}}d\chi\;e^{-\tau(\chi)}\,n_\mathrm{e}(\chi)\,a(\chi)\int\int \frac{d^3\mathbf{k_1}d^3\mathbf{k_2}}{(2\pi)^6}\tilde{u}(\mathbf{k_{1}})\tilde{u}(\mathbf{k_{2}})\nonumber\\
		&\hspace{1.5cm} \sum_{{L, M}}i^L Y_{LM}^{*}(\hat{\mathbf{k}})\,j_{L}(k\chi)\sum_{p_1,p_2}
		\left(\begin{array}{ccc}
			1& 1 & 2\\ 
			p_1& p_2 & -\lambda
		\end{array}\right)Y_{1p_1}^{*}(\mathbf{{\hat{k}}_1})Y_{1p_2}^{*}(\mathbf{{\hat{k}_2}})\;A^{\lambda L M}_{\ell m},
	\end{align}
	where
	\begin{align}
		A^{\lambda L M}_{\ell m}=\sqrt{\frac{5(2L+1)(2\ell +1)}{4\pi}}(-1)^{(m)}\:
		\left(\begin{array}{ccc}
			L& 2 & \ell\\ 
			0& -2 & 2
		\end{array}\right)
		\left(\begin{array}{ccc}
			L& 2& \ell\\ 
			M& \lambda & -m
		\end{array}\right).
	\end{align}
	Using eq.(\ref{e_b_coeff}) we can now obtain the E mode and B mode  harmonic coefficients. To simplify further, we use the following properties of Wigner 3j symbols,
	\bea
		\left(\begin{array}{ccc}
		l_1& l_2& l_3\\ 
		-m_1& -m_2 & m_3
	\end{array}\right)=(-1)^{(l_1+l_2+l_3)}	
\left(\begin{array}{ccc}
	l_1& l_2& l_3\\ 
	m_1& m_2 & -m_3
\end{array}\right).
	\eea
	After some algebraic manipulations, (see Appendix \ref{App:EB_coeff_power} for more details) we get the following expression for the E and B mode harmonic coefficients,
	\begin{align}
		\label{e_harmonic_coeff}
		\hspace{-0.3cm}e_{\ell m}=&\frac{1}{2}(4\pi)\left(\frac{4\pi}{3}\right)^2\sqrt{\frac{3}{2\pi}}\frac{\sqrt{6}\sigma_{\mathrm{T}}}{10}\sum_{\lambda =-2}^{2}(-1)^\lambda \int_{0}^{\chi_{i}}d\chi\;e^{-\tau(\chi)}\,n_\mathrm{e}(\chi)\,a(\chi)\int \int \frac{d^3\mathbf{k_1}d^3\mathbf{k_2}}{(2\pi)^6} \tilde{u}(\mathbf{k_{1}})\tilde{u}(\mathbf{k_{2}})\nonumber\\
		& \sum_{{L, M}}i^L Y_{L M}^{*}(\hat{\mathbf{k}})\,j_{L}(k\chi)\sum_{p_1,p_2}
		\left(\begin{array}{ccc}
			1& 1 & 2\\ 
			p_1& p_2 & -\lambda
		\end{array}\right)Y_{1p_1}^{*}(\mathbf{{\hat{k}}_1})Y_{1p_2}^{*}(\mathbf{{\hat{k}_2}})\;A^{\lambda L M}_{\ell m}\left(1+(-1)^{(L+\ell)}\right)
	\end{align}
	and 
	\begin{align}
		\label{b_harmonic_coeff}
\hspace{-0.3cm}	b_{\ell m}=&-\frac{i}{2}(4\pi)\left(\frac{4\pi}{3}\right)^2\sqrt{\frac{3}{2\pi}}\frac{\sqrt{6}\sigma_{\mathrm{T}}}{10}\sum_{\lambda =-2}^{2}(-1)^\lambda \int_{0}^{\chi_{i}}d\chi\;e^{-\tau(\chi)}\,n_\mathrm{e}(\chi)\,a(\chi)\int \int \frac{d^3\mathbf{k_1}d^3\mathbf{k_2}}{(2\pi)^6} \tilde{u}(\mathbf{k_{1}})\tilde{u}(\mathbf{k_{2}})\nonumber\\
		& \sum_{{L, M}}i^L Y_{L M}^{*}(\hat{\mathbf{k}})\,j_{L}(k\chi)\sum_{p_1,p_2}
		\left(\begin{array}{ccc}
			1& 1 & 2\\ 
			p_1& p_2 & -\lambda
		\end{array}\right)Y_{1p_1}^{*}(\mathbf{{\hat{k}}_1})Y_{1p_2}^{*}(\mathbf{{\hat{k}_2}})\;A^{\lambda L M}_{\ell m}\left(1-(-1)^{(L+\ell)}\right).
	\end{align}
     
	\section{Power spectrum of E and B modes \label{sec_EB_pow}}
	Taking the ensemble average, denoted by the angular brackets $\langle\:\:\rangle$, gives us the auto-spectra of the E mode $\langle e_{\ell m}e^{*}_{\ell' m'}\rangle$ and the B mode $\langle b_{\ell m}b^{*}_{\ell' m'}\rangle$ polarisation. We have explicitly checked that they are diagonal in the harmonic space, i.e. 
	\bea
	\langle e_{\ell m}e^{*}_{\ell' m'}\rangle=C^{EE}_{\ell}\;\delta_{\ell,\ell'}\;\delta_{m,m'} \hspace{0.4in} \mathrm{and}\hspace{0.4in} \langle b_{\ell m}b^{*}_{\ell' m'}\rangle=C^{BB}_{\ell}\;\delta_{\ell,\ell'}\;\delta_{m,m'},
	\eea 
	as expected from statistical homogeneity and isotropy. Thus, we a priori choose $\ell=\ell'$ and $m=m'=0$ for our numerical calculations. From eq.(\ref{e_harmonic_coeff}) and eq.(\ref{b_harmonic_coeff}) we get
	\begin{align}
		\label{cl_ee_1}
		C^{EE}_{\ell}=&	\frac{T^{2}_{\mathrm{CMB}}}{4}\left[(4\pi)\left(\frac{4\pi}{3}\right)^2\sqrt{\frac{3}{2\pi}}\frac{\sqrt{6}\sigma_{\mathrm{T}}}{10}\right]^2\sum_{\lambda,\lambda' =-2}^{2}(-1)^{(\lambda+\lambda')}\int_{0}^{\chi_{i}}d\chi\;e^{-\tau(\chi)}\, a(\chi)n_\mathrm{e}(\chi)\nonumber\\
		&\int_{0}^{\chi_{i}}d\chi'\;e^{-\tau(\chi')} a(\chi')n_\mathrm{e}(\chi')\int \int \frac{d^3\mathbf{k_1}d^3\mathbf{k_2}}{(2\pi)^6} \int \int \frac{d^3\mathbf{k_1'}d^3\mathbf{k_2'}}{(2\pi)^6}  \,\Big\langle \tilde{u}(\mathbf{k_{1}})\tilde{u}(\mathbf{k_{2}})\tilde{u}^{*}(\mathbf{k_{1}'})\tilde{u}^{*}(\mathbf{k_{2}'})\Big\rangle \nonumber\\
		&\sum_{{L, M}\atop{L',M'}}i^{(L-L')}Y_{L M}^{*}(\hat{\mathbf{k}})Y_{L' M'}(\hat{\mathbf{k}}')\;j_{L}(k\chi)\,j_{L'}(k'\chi')\sum_{{p_1,p_2}\atop{p_1',p_2'}}
		\left(\begin{array}{ccc}
			1& 1 & 2\\ 
			p_1& p_2 & -\lambda
		\end{array}\right)
	\left(\begin{array}{ccc}
		1& 1 & 2\\ 
		p_1'& p_2' & -\lambda'
	\end{array}\right)\nonumber\\
		&Y_{1p_1}^{*}(\mathbf{{\hat{k}}_1})Y_{1p_2}^{*}(\mathbf{{\hat{k}_2}})Y_{1p_1'}(\mathbf{{\hat{k}}_1'})Y_{1p_2'}(\mathbf{{\hat{k}_2'}})A^{\lambda L M}_{\ell m}A^{\lambda' L' M'}_{\ell m}\left(1+(-1)^{(L+\ell)}\right)\left(1+(-1)^{(L'+\ell)}\right)
	\end{align}
	and 
	\begin{align}
		\label{cl_bb_1}
		C^{BB}_{\ell}=&	\frac{T^{2}_{\mathrm{CMB}}}{4}\left[(4\pi)\left(\frac{4\pi}{3}\right)^2\sqrt{\frac{3}{2\pi}}\frac{\sqrt{6}\sigma_{\mathrm{T}}}{10}\right]^2\sum_{\lambda,\lambda' =-2}^{2}(-1)^{(\lambda+\lambda')}\int_{0}^{\chi_{i}}d\chi\;e^{-\tau(\chi)}\, a(\chi)n_\mathrm{e}(\chi)\nonumber\\
		&\int_{0}^{\chi_{i}}d\chi'\;e^{-\tau(\chi')}a(\chi')n_\mathrm{e}(\chi')\int \int \frac{d^3\mathbf{k_1}d^3\mathbf{k_2}}{(2\pi)^6}\int \int \frac{d^3\mathbf{k_1'}d^3\mathbf{k_2'}}{(2\pi)^6}  \,\Big\langle \tilde{u}(\mathbf{k_{1}})\tilde{u}(\mathbf{k_{2}})\tilde{u}^{*}(\mathbf{k_{1}'})\tilde{u}^{*}(\mathbf{k_{2}'})\Big\rangle \nonumber\\
		&\sum_{{L, M}\atop{L',M'}}i^{(L-L')}Y_{L M}^{*}(\hat{\mathbf{k}})Y_{L' M'}(\hat{\mathbf{k}}')\;j_{L}(k\chi)\,j_{L'}(k'\chi')\sum_{{p_1,p_2}\atop{p_1',p_2'}}
			\left(\begin{array}{ccc}
			1& 1 & 2\\ 
			p_1& p_2 & -\lambda
		\end{array}\right)
		\left(\begin{array}{ccc}
			1& 1 & 2\\ 
			p_1'& p_2' & -\lambda'
		\end{array}\right)
	\nonumber\\
		&Y_{1p_1}^{*}(\mathbf{{\hat{k}}_1})Y_{1p_2}^{*}(\mathbf{{\hat{k}_2}})Y_{1p_1'}(\mathbf{{\hat{k}}_1'})Y_{1p_2'}(\mathbf{{\hat{k}_2'}})A^{\lambda L M}_{\ell m}A^{\lambda' L' M'}_{\ell m}\left(1-(-1)^{(L+\ell)}\right)\left(1-(-1)^{(L'+\ell)}\right),
	\end{align}
	where we have multiplied the expressions by $T^{2}_{\mathrm{CMB}}$ ( $T_{\mathrm{CMB}}=2.725$ K), to give the results in temperature units. 
	We need to calculate the ensemble average over the velocity potentials. These are Gaussian random fields. We use the Isserlis theorem to break the 4 point function,
	\begin{align}
		\label{corr_expansion_0}
		\Big\langle \tilde{u}(\mathbf{k_{1}})\tilde{u}(\mathbf{k_{2}})\tilde{u}^{*}(\mathbf{k_{1}'})\tilde{u}^{*}(\mathbf{k_{2}'})\Big\rangle=&\Big\langle \tilde{u}(\mathbf{k_{1}})\tilde{u}(\mathbf{k_{2}})\Big\rangle\Big\langle \tilde{u}^{*}(\mathbf{k_{1}'})\tilde{u}^{*}(\mathbf{k_{2}'})\Big\rangle+\Big\langle \tilde{u}(\mathbf{k_{1}}) u^{*}(\mathbf{k_{1}'})\Big\rangle\Big\langle \tilde{u}(\mathbf{k_{2}})u^{*}(\mathbf{k_{2}'})\Big\rangle\nonumber\\
		&+\Big\langle \tilde{u}(\mathbf{k_{1}}) u^{*}(\mathbf{k_{2}'})\Big\rangle\Big\langle \tilde{u}(\mathbf{k_{2}})u^{*}(\mathbf{k_{1}'})\Big\rangle.
	\end{align}
	The ensemble average over the velocity potentials is given by
	\begin{align}
	\label{ensemble_def}
	\Big\langle\tilde{u}(\mathbf{k})\tilde{u}^{*}(\mathbf{k'})\Big\rangle=(2\pi)^3P_{uu}(k)\,\delta(\mathbf{k}-\mathbf{k'})\hspace{0.2in}\mathrm{and}\hspace{0.2in}
	\Big\langle\tilde{u}(\mathbf{k})\tilde{u}(\mathbf{k'})\Big\rangle=(2\pi)^3P_{uu}(k)\,\delta(\mathbf{k}+\mathbf{k'}),
	\end{align}
	where 
	\bea
	P_{uu}(k)=\frac{\left(aH(a)f(a)\right)^2}{k^2}P_{L}(k,a,a'),
	\eea 
	$P_{L}$ is the linear matter power spectrum and f is the growth rate \cite{dodelson2003modern},
	\bea
	 \left(f= \frac{d\,\ln D_+(a)}{d\,\ln a}\simeq \left[\Omega_m(a)\right]^{0.55}\right).
	 \eea
	 We have used the linear matter power given by Colossus \cite{diemer2018colossus} which uses the model given in \cite{1998Eisenstein_Hu} to calculate the transfer function. It can be easily shown on doing the angular integrals, that the first term in eq.(\ref{corr_expansion_0}) does not contribute to the power spectrum (see Appendix \ref{App:4point}). The contributions from the second and the third terms are equal as eq.(\ref{cl_ee_1}) and eq.(\ref{cl_bb_1}) are symmetric under the exchange of $\mathbf{k_1}$ and   $\mathbf{k_2}$. Hence, we only need to consider one of the terms, $\Big\langle \tilde{u}(\mathbf{k_{1}}) u^{*}(\mathbf{k_{1}'})\Big\rangle\Big\langle \tilde{u}(\mathbf{k_{2}})u^{*}(\mathbf{k_{2}'})\Big\rangle$. Using eq.(\ref{ensemble_def}), we get (in temperature units),
	
	\begin{align}
		\label{cl_ee_2}
		C^{EE}_{\ell}=&	\frac{T^{2}_{\mathrm{CMB}}}{2}\left[(4\pi)\left(\frac{4\pi}{3}\right)^2\sqrt{\frac{3}{2\pi}}\frac{\sqrt{6}\sigma_{\mathrm{T}}}{10}\right]^2\sum_{\lambda,\lambda' =-2}^{2}(-1)^{(\lambda+\lambda')}\int_{0}^{\chi_{i}}d\chi\;e^{-\tau(\chi)}\, a(\chi) n_\mathrm{e}(\chi)\nonumber\\
		&\int_{0}^{\chi_{i}}d\chi'\;e^{-\tau(\chi')}a(\chi')n_\mathrm{e}(\chi')\sum_{{L, M}\atop{L',M'}}\sum_{{p_1,p_2}\atop{p_1',p_2'}}i^{(L-L')}	
		\left(\begin{array}{ccc}
			1& 1 & 2\\ 
			p_1& p_2 & -\lambda
		\end{array}\right)
		\left(\begin{array}{ccc}
			1& 1 & 2\\ 
			p_1'& p_2' & -\lambda'
		\end{array}\right)\nonumber\\
		&\int \int \frac{dk_1k^2_{1}\,dk_2k^2_{2}}{(2\pi)^6}P_{uu}(k_1)P_{uu}(k_2)j_{L}(k\chi)j_{L'}(k'\chi')\int d\Omega_{\mathbf{k_1}}\int d\Omega_{\mathbf{k_2}}\;Y_{L M}^{*}(\hat{\mathbf{k}})Y_{L'M'}(\hat{\mathbf{k}})\nonumber\\
		&Y_{1p_1}^{*}(\mathbf{{\hat{k}}_1})Y_{1p_2}^{*}(\mathbf{{\hat{k}_2}})Y_{1p_1'}(\mathbf{{\hat{k}}_1})Y_{1p_2'}(\mathbf{{\hat{k}_2}})A^{\lambda L M}_{\ell m}A^{\lambda' L' M'}_{\ell m}\left(1+(-1)^{(L+\ell)}\right)\left(1+(-1)^{(L'+\ell)}\right)
	\end{align}
	and 
	\begin{align}
		\label{cl_bb_2}
		C^{BB}_{\ell}=&	\frac{T^{2}_{\mathrm{CMB}}}{2}\left[(4\pi)\left(\frac{4\pi}{3}\right)^2\sqrt{\frac{3}{2\pi}}\frac{\sqrt{6}\sigma_{\mathrm{T}}}{10}\right]^2\sum_{\lambda,\lambda' =-2}^{2}(-1)^{(\lambda+\lambda')}\int_{0}^{\chi_{i}}d\chi\;e^{-\tau(\chi)}\, a(\chi) n_\mathrm{e}(\chi)\nonumber\\
		&\int_{0}^{\chi_{i}}d\chi'\;e^{-\tau(\chi')}a(\chi')n_\mathrm{e}(\chi')\sum_{{L, M}\atop{L',M'}}\sum_{{p_1,p_2}\atop{p_1',p_2'}}i^{(L-L')}	
		\left(\begin{array}{ccc}
			1& 1 & 2\\ 
			p_1& p_2 & -\lambda
		\end{array}\right)
		\left(\begin{array}{ccc}
			1& 1 & 2\\ 
			p_1'& p_2' & -\lambda'
		\end{array}\right)\nonumber\\
		&\int \int \frac{dk_1k^2_{1}\,dk_2k^2_{2}}{(2\pi)^6}P_{uu}(k_1)P_{uu}(k_2)j_{L}(k\chi)j_{L'}(k'\chi')\int d\Omega_{\mathbf{k_1}}\int d\Omega_{\mathbf{k_2}}\;Y_{L M}^{*}(\hat{\mathbf{k}})Y_{L'M'}(\hat{\mathbf{k}})\nonumber\\
		&Y_{1p_1}^{*}(\mathbf{{\hat{k}}_1})Y_{1p_2}^{*}(\mathbf{{\hat{k}_2}})Y_{1p_1'}(\mathbf{{\hat{k}}_1})Y_{1p_2'}(\mathbf{{\hat{k}_2}})A^{\lambda L M}_{\ell m}A^{\lambda' L' M'}_{\ell m}\left(1-(-1)^{(L+\ell)}\right)\left(1-(-1)^{(L'+\ell)}\right).
	\end{align}
	We should note that the above expressions for the angular power spectra is derived  by taking $\left(\mathbf{v}\cdot\mathbf{\hat{n}'}\right)^2$ as the source term from eq.(\ref{k_quadrupole}) in eq.(\ref{pol_final}) with unit normalisation. The actual signal depends on the frequency of observation with spectra given by eq.(\ref{k_quadrupole}). 
	
	From component separation perspective, we want to decompose the spectrum into a differential blackbody part $g(x)$ and y-type distortion part, $y(x)$. In particular, so that we are not affected by the cosmic variance of the primary CMB E-modes and the lensing B-modes, we want a strategy which will eliminate the blackbody part while preserving the y-distortion part. In the Rayleigh-Jeans (small frequency)  limit, we see that the expressions of y(x) in eq.(\ref{temp_quadpole}) becomes equal to $-2$, i.e. $\lim_{x\rightarrow0} y(x)=-2$, which cancels the factor of 1/2 multiplying $y(x)$ in eq.(\ref{k_quadrupole}). Therefore eq.(\ref{cl_ee_2}) and eq.(\ref{cl_bb_2}) give the y-type E and B mode power spectra in the RJ limit in temperature units.  For the blackbody part the amplitude is equal to $2\left(\mathbf{v}\cdot\mathbf{\hat{n}'}\right)^2$. Thus, the blackbody power spectra are actually 4 times the above expressions in temperature units.
	\bea
	C^{BB \;\mathrm{y-type}}_{\ell}\Big|_{\mathrm{RJ}}=C^{BB}_{\ell} \;\;\mathrm{and} \hspace{1cm} C^{EE\; \mathrm{y-type}}_{\ell}\Big|_{\mathrm{RJ}}= C^{EE}_{\ell}.
	\eea
	and just for the blackbody part for both E and B modes we have,
	\bea
	C^{\; \mathrm{y-type}}_\ell\Big|_{\mathrm{RJ}}=\frac{1}{4}C^{\mathrm{Blackbody}}_{\ell}.
	\eea
	We can now integrate these expressions numerically to obtain the final results.  For the integrals over radial $k_1$ and $k_2$ modes, we integrate from $10^{-5}\;\mathrm{Mpc^{-1}}$ to $1.5\;\mathrm{Mpc^{-1}}$ in comoving coordinates. The line of sight integration over comoving distances $\chi$ and $\chi'$ are from $z=20$ before the reionisation starts until $z=0$. We consider two models for the reionisation history. A redshift symmetric model defined by hyperbolic tangent function \cite{lewis2000efficient, heinrich2017complete} and a redshift asymmetric model \cite{DouspisAsymmetric}. The ionisation fraction is defined as the ratio between the electron number density and the total hydrogen number density at that redshift, $\mathrm{X_e}(z)=\left(\frac{n_\mathrm{e}(z)}{n_\mathrm{H}(z)}\right)$, where $n_\mathrm{H}(z)=n_\mathrm{H}(0)\left(1+z\right)^3$, $n_\mathrm{H}(0)$ is the hydrogen number density at $z=0$ assuming primordial abundance. We assume the first helium reionisation to proceed identically. Thus, for the symmetric model the ionisation fraction is given as
	\begin{align}
		\label{reion_history}
		\mathrm{X_e}^{\mathrm{Sym}}(z)=\left[\frac{(1+f)}{2}\left\{1+\tanh\left(\frac{q_{\mathrm{re}}-q}{\Delta q_{\mathrm{re}}}\right)\right\}+\frac{f}{2}\left\{1+\tanh\left(\frac{q^{\mathrm{HeII}}_{\mathrm{re}}-q}{\Delta q^{\mathrm{HeII}}_{\mathrm{re}}}\right)\right\}\right]
	\end{align}
	and for asymmetric case,
	\begin{align}
		\label{reion_history_asym}
		\mathrm{X_e}^{\mathrm{Asym}}(z)=\left[\left\{{(1+f) \hspace{2.9cm}z<z_{\mathrm{end}}\atop (1+f)\left(\frac{z_{\mathrm{early}}-z}{z_{\mathrm{early}}-z_{\mathrm{end}}}\right)^\alpha\hspace{0.5cm}z>z_{\mathrm{end}}}\right\}
		+\frac{f}{2}\left\{1+\tanh\left(\frac{q^{\mathrm{HeII}}_{\mathrm{re}}-q}{\Delta q^{\mathrm{HeII}}_{\mathrm{re}}}\right)\right\}\right],
	\end{align}
	where $q(z)=\left(1+z\right)^{1.5}$, $q_{\mathrm{re}}=q(z_{\mathrm{re}})$, $\Delta q_{\mathrm{re}}=1.5(\sqrt{1+z_{\mathrm{re}}})\beta_{\mathrm{re}}$, and $f=\left(\frac{\mathrm{m_H}}{\mathrm{m_{He}}}\frac{\mathrm{X_{He}}}{1-\mathrm{X_{He}}}\right)\simeq0.079$. The central redshift of reionisation is given by $z_{\mathrm{re}}$ and $\beta_{\mathrm{re}}$ is a parameter characterising how fast reionisation happens. We also define the duration of reionisation as $\Delta z_{\mathrm{re}}=z_{10\%}-z_{99\%}$, where $z_{x \%} $ is the redshift when $\frac{X_e}{(1+f)}=\frac{x}{100}$, i.e. the hydrogen is $x \%$ ionised. For the $\mathrm{2^{nd}}$ reionisation of Helium, we always use a hyperbolic tangent function. It happens at redshift $\sim\, 3$. We have chosen $\beta^{\mathrm{HeII}}_{\mathrm{re}}=3.5$ and $z^{\mathrm{HeII}}_{\mathrm{re}}=0.5$. We note that $\mathrm{HeII}$ reionisation gives a negligible contribution to the signal, but we have included it for completeness. We have also fixed $z_{\mathrm{early}}=20$ and $z_{\mathrm{end}}=6$ in the asymmetric model. In this case, the exponent $\alpha$ determines the rapidity with which reionisation takes place. The reionisation histories for different reionisation parameters are shown in figure \ref{fig:reion_history}  in appendix \ref{App:Reion_history}. 
	
		\begin{table}
		\centering
		\begin{subtable}{0.6\textwidth}
			\vspace{0.0cm}
			\hspace{-0.05cm}
			\begin{minipage}[c]{0.9\textwidth}
\begin{tabular}{|ccc|}
	\hline
	\multicolumn{3}{|c|}{At $\beta_{\mathrm{re}}= 0.5$}                                                                                                                                                                                                 \\ \hline
	\multicolumn{1}{|c|}{\begin{tabular}[c]{@{}c@{}}Central \\Redshift $(z_{\mathrm{re}})$\end{tabular}} & \multicolumn{1}{c|}{\begin{tabular}[c]{@{}c@{}} Duration\\ $\left(z_{10\%}-z_{99\%}\right)$\end{tabular}} & Optical depth $\tau$ \\ \hline
	\multicolumn{1}{|c|}{7.5}                                                                                  & \multicolumn{1}{c|}{1.7323}                                                                                     & 0.05534              \\ \hline
	\multicolumn{1}{|c|}{8.5}                                                                                  & \multicolumn{1}{c|}{1.7283}                                                                                     & 0.06587              \\ \hline
	\multicolumn{1}{|c|}{9.5}                                                                                  & \multicolumn{1}{c|}{1.7251}                                                                                     & 0.07697              \\ \hline
\end{tabular}
			\caption{For different central redshift, fixing $\beta_{\mathrm{re}}= 0.5$.}
			\label{tab: table1 }
		\end{minipage}
		\end{subtable}%
		\begin{subtable}{0.5\textwidth}
			\vspace{0.4cm}
			\hspace{-0.49cm}
			\begin{minipage}[c]{0.87\textwidth}
\begin{tabular}{|ccc|}
	\hline
	\multicolumn{3}{|c|}{At central redshift $z_{\mathrm{re}}=8.5$}                                                                                                               \\ \hline
	\multicolumn{1}{|c|}{$\beta_{\mathrm{re}}$} & \multicolumn{1}{c|}{\begin{tabular}[c]{@{}c@{}}Duration\\ $\left(z_{10\%}-z_{99\%}\right)$\end{tabular}} & Optical depth $\tau$ \\ \hline
	\multicolumn{1}{|c|}{0.1}                   & \multicolumn{1}{c|}{0.340}                                                                               & 0.065871            \\ \hline
	\multicolumn{1}{|c|}{0.5}                   & \multicolumn{1}{c|}{1.728}                                                                               & 0.065872           \\ \hline
	\multicolumn{1}{|c|}{1.3}                   & \multicolumn{1}{c|}{4.670}                                                                               & 0.065875             \\ \hline
\end{tabular}
			\caption{For different width of reionisation, fixing central redshift at $z_{\mathrm{re}}=8.5$.}
			\label{tab: table2 }
		\end{minipage}
		\end{subtable}
		\begin{subtable}{0.5\textwidth}
			\vspace{0.4cm}
			\hspace{0.0in}
			\begin{minipage}[c]{0.85\textwidth}
\begin{tabular}{|ccc|}
	\hline
	\multicolumn{3}{|c|}{At $\mathrm{z_{early}}=20$ and $\mathrm{z_{end}}=6$}                                                                                        \\ \hline
	\multicolumn{1}{|c|}{$\alpha$} & \multicolumn{1}{c|}{\begin{tabular}[c]{@{}c@{}}Duration\\ $\left(z_{10\%}-z_{99\%}\right)$\end{tabular}} & Optical depth $\tau$ \\ \hline
	\multicolumn{1}{|c|}{3}        & \multicolumn{1}{c|}{7.454}                                                                               & 0.07891              \\ \hline
	\multicolumn{1}{|c|}{5}        & \multicolumn{1}{c|}{5.138}                                                                               & 0.06516              \\ \hline
	\multicolumn{1}{|c|}{12}       & \multicolumn{1}{c|}{2.432}                                                                               & 0.05137              \\ \hline
\end{tabular}
			\caption{For different rapidity parameter $\alpha$, fixing $\mathrm{z_{early}}=20$ and $\mathrm{z_{end}}=6$.}
			\label{tab: table3 }
		\end{minipage}
		\end{subtable}
		\caption{The total optical depth and duration of reionisation for different reionisation model parameters.}
	\end{table}
	\section{Results \label{sec_result}}
	We performed the integrals  in eq.(\ref{cl_bb_2}) and eq.(\ref{cl_ee_2}) numerically using multidimensional Adaptive Monte Carlo integration, Vegas algorithm \cite{lepage1978new}. We have checked the saturation of our numerical results by increasing the Monte Carlo steps until the result converges and the standard deviation is at least an order of magnitude less than the mean value. The angular power spectrum of the y-type E and B modes in the RJ limit of the spectrum in temperature units, for a symmetric reionisation history with central redshift $z_{\mathrm{re}}= 8.5$ and $\beta_{\mathrm{re}}=0.5$, is plotted in figure \ref{fig:summary plot}. The primordial B modes at tensor to scalar ratio $r$ of $10^{-4}$  and $10^{-5}$, the PRISM sensitivity curve, and the PICO sensitivity curve are also plotted for comparison. This figure shows that detecting the y-type E and B modes will be challenging. The blackbody component of the pkSZ B modes is 4 times larger compared to the y-type power spectrum in figure \ref{fig:summary plot} and will start to be an important foreground for the detection of primordial B modes at tensor to scalar ratio $r\lesssim 3\times10^{-5}$.

	We also study the sensitivity of the pkSZ effect to the reionisation history. In table \ref{tab: table1 }, table \ref{tab: table2 }, and table \ref{tab: table3 }, we show the values for different reionisation parameters used in the analysis and the corresponding total Thomson optical depths as defined by eq.(\ref{tau_chi}). 
	The power spectrum for these different reionisation histories are shown in figure \ref{fig:effect_redshift}, figure \ref{fig:effect_redshift_asym} and figure \ref{fig:effect_width}. For all the cases of symmetric reionisation, we have compared the power spectrum curves with the fiducial case, having parameters $ \beta_{\mathrm{re}}=8.5$ and $ \beta_{\mathrm{re}}=0.5$. 
	\subsection{Dependence on optical depth  \label{subsec_central_redshift}}
The main effect of changing the central redshift of reionisation in the symmetric model is the obvious change in the total optical depth as shown in table \ref{tab: table1 }.
\begin{figure}
			\hspace{-0.65cm}
		\begin{subfigure}{0.5\textwidth}
			\centering
			\includegraphics[width=1.095\linewidth]{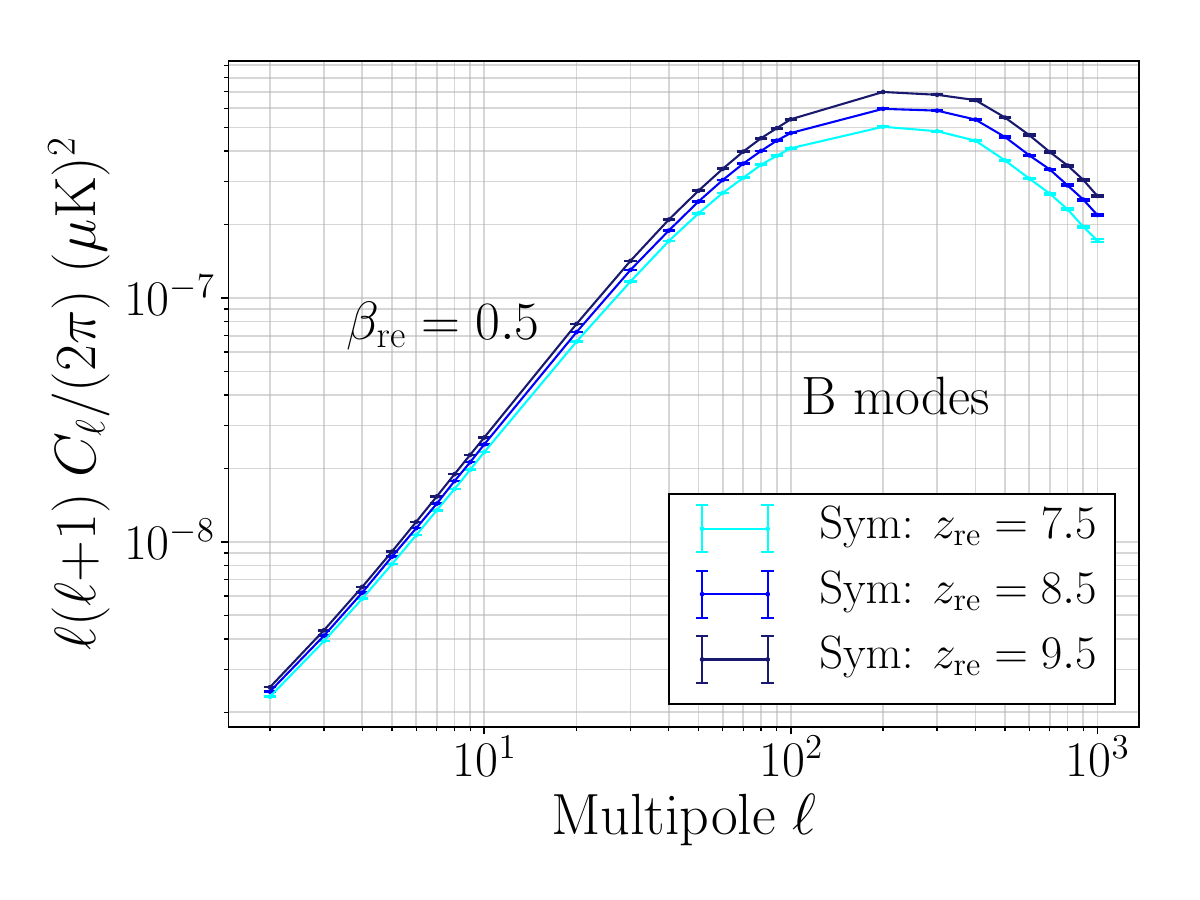}
			\caption{B modes}
			\label{fig:bmodes_sym}
		\end{subfigure}%
	\hspace{0.18cm}
		\begin{subfigure}{0.5\textwidth}
			\centering
			\includegraphics[width=1.095\linewidth]{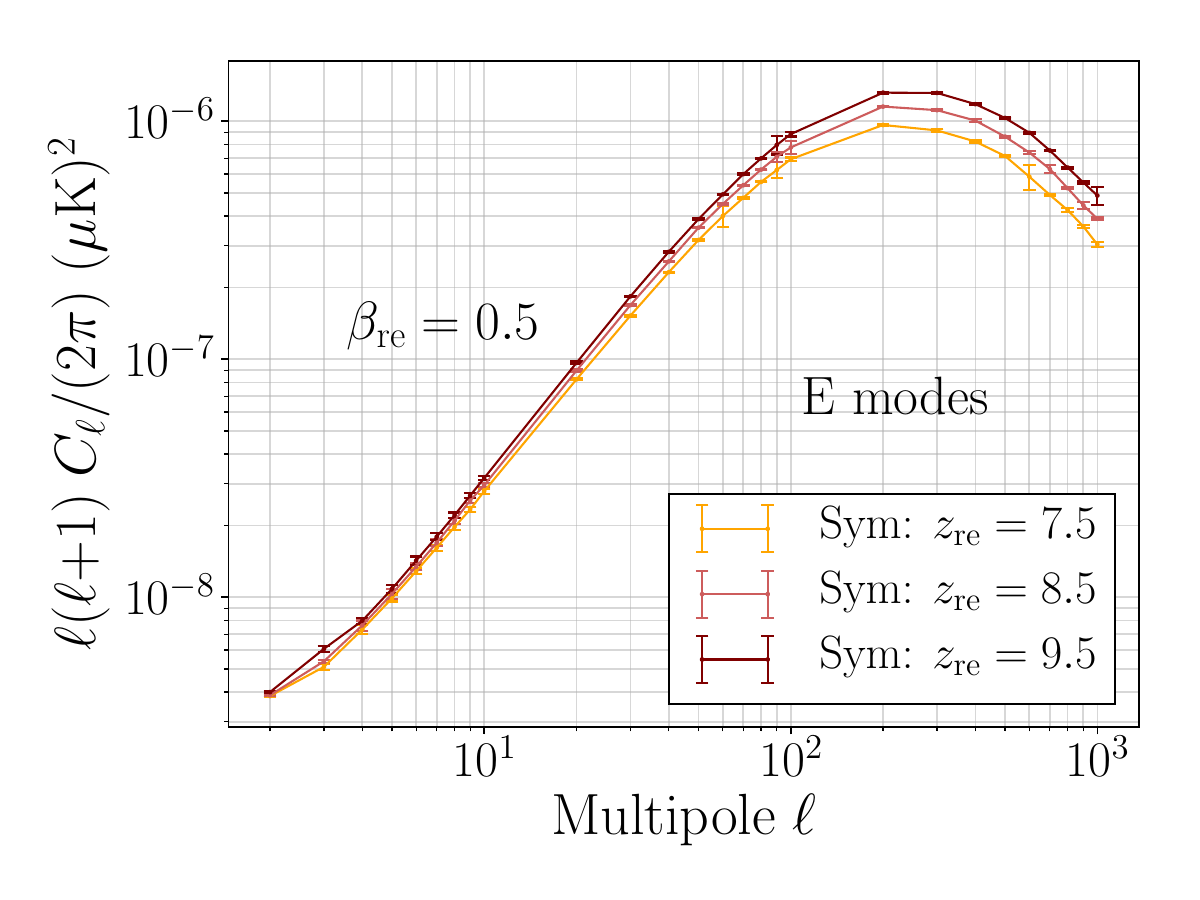}
			\caption{E modes}
			\label{fig:emode_sym}
		\end{subfigure}
		\vfill
			\hspace{-0.5cm}
	\begin{subfigure}{0.5\textwidth}
		\centering
		\includegraphics[width=1.075\linewidth]{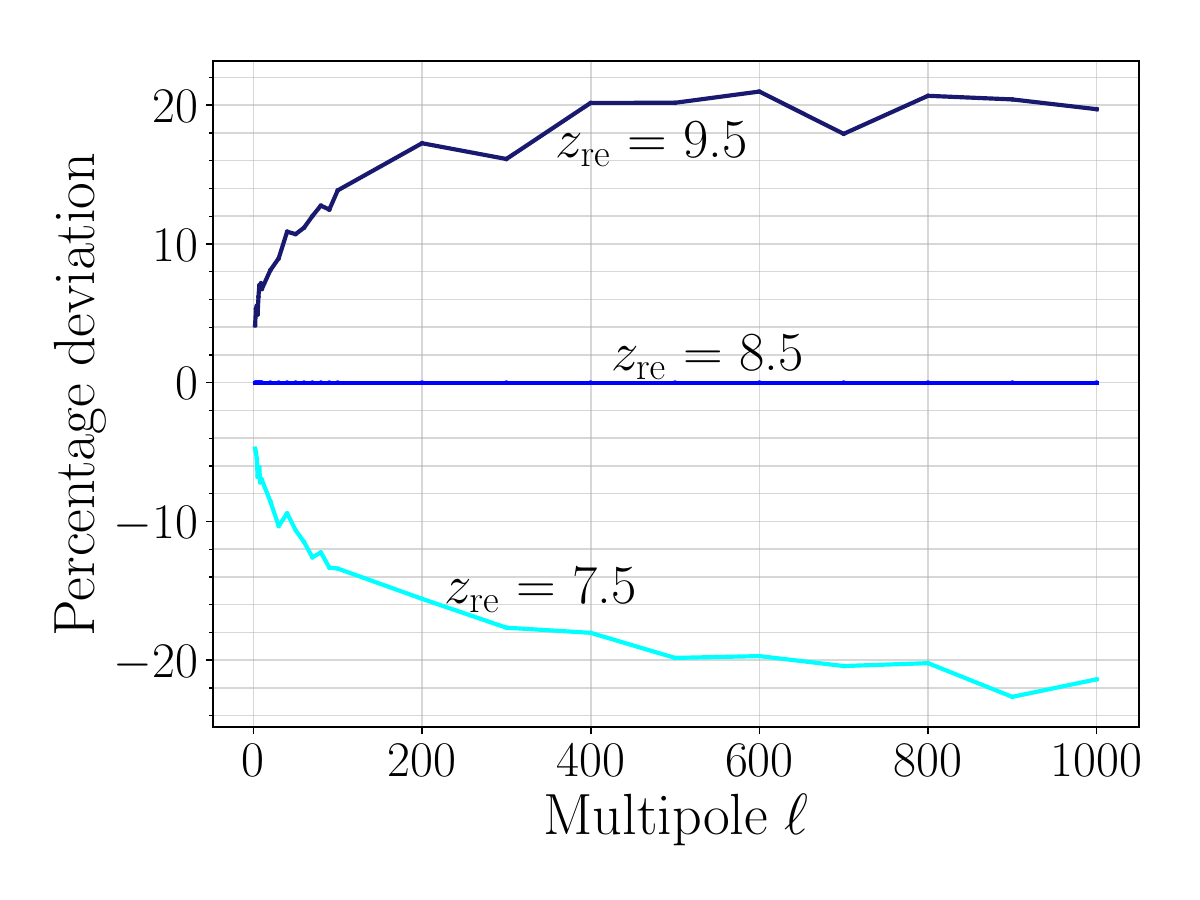}
		\caption{Percentage change in B modes}
		\label{fig:bmode_frac_sym}
	\end{subfigure}	
	\hspace{0.0cm}
	\begin{subfigure}{0.5\textwidth}
		\centering
		\includegraphics[width=1.075\linewidth]{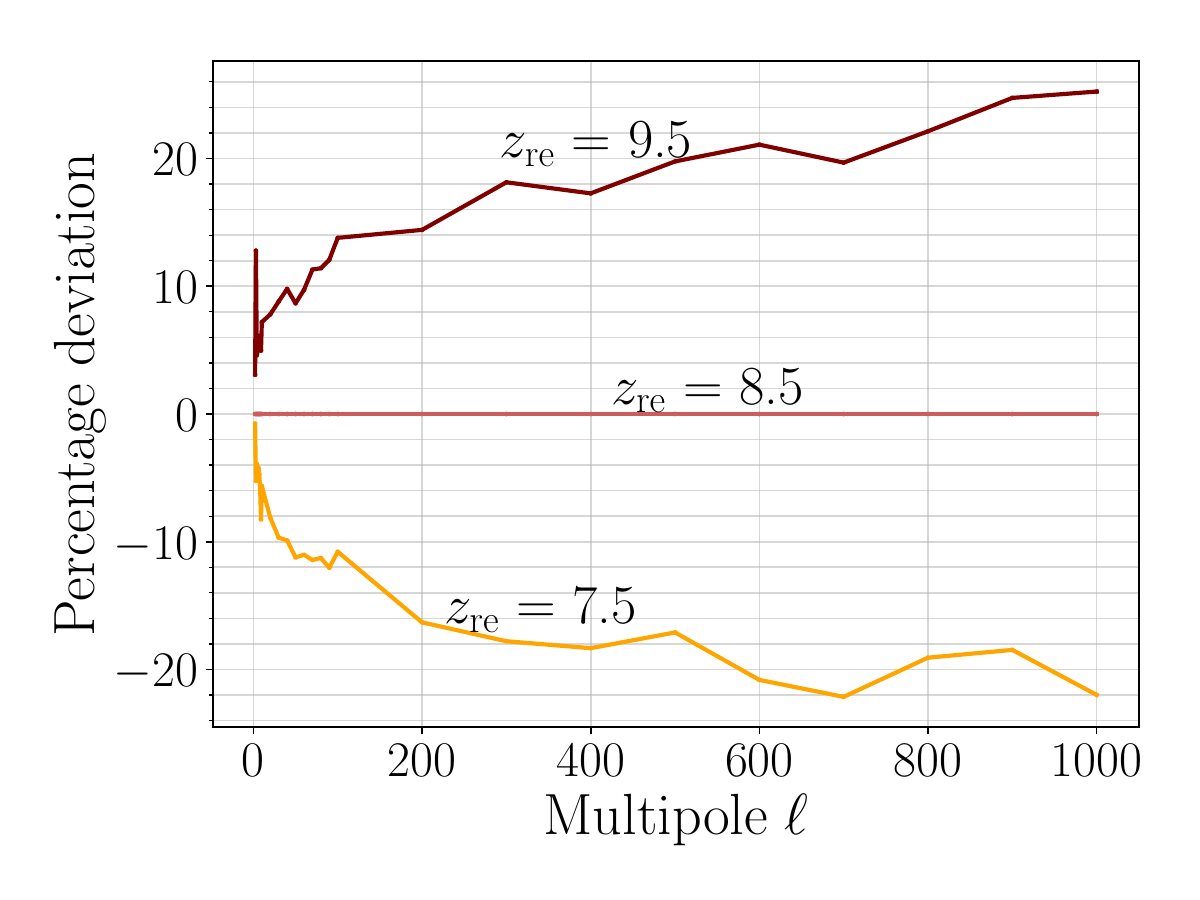}
		\caption{Percentage change in E modes}
		\label{fig:emode_frac_sym}
	\end{subfigure}	
		\caption{The angular power spectra for central redshift $z_{\mathrm{re}}$ at 7.5, 8.5, and 9.5, keeping $\beta_{\mathrm{re}}=0.5$. The optical depth increases as $z_{\mathrm{re}}$ increases, resulting in an increase in the angular power spectrum.}
		\label{fig:effect_redshift}
	\end{figure}
	\begin{figure}
			\hspace{-0.5cm}
	\begin{subfigure}{0.5\textwidth}
		\centering
		\includegraphics[width=1.08\linewidth]{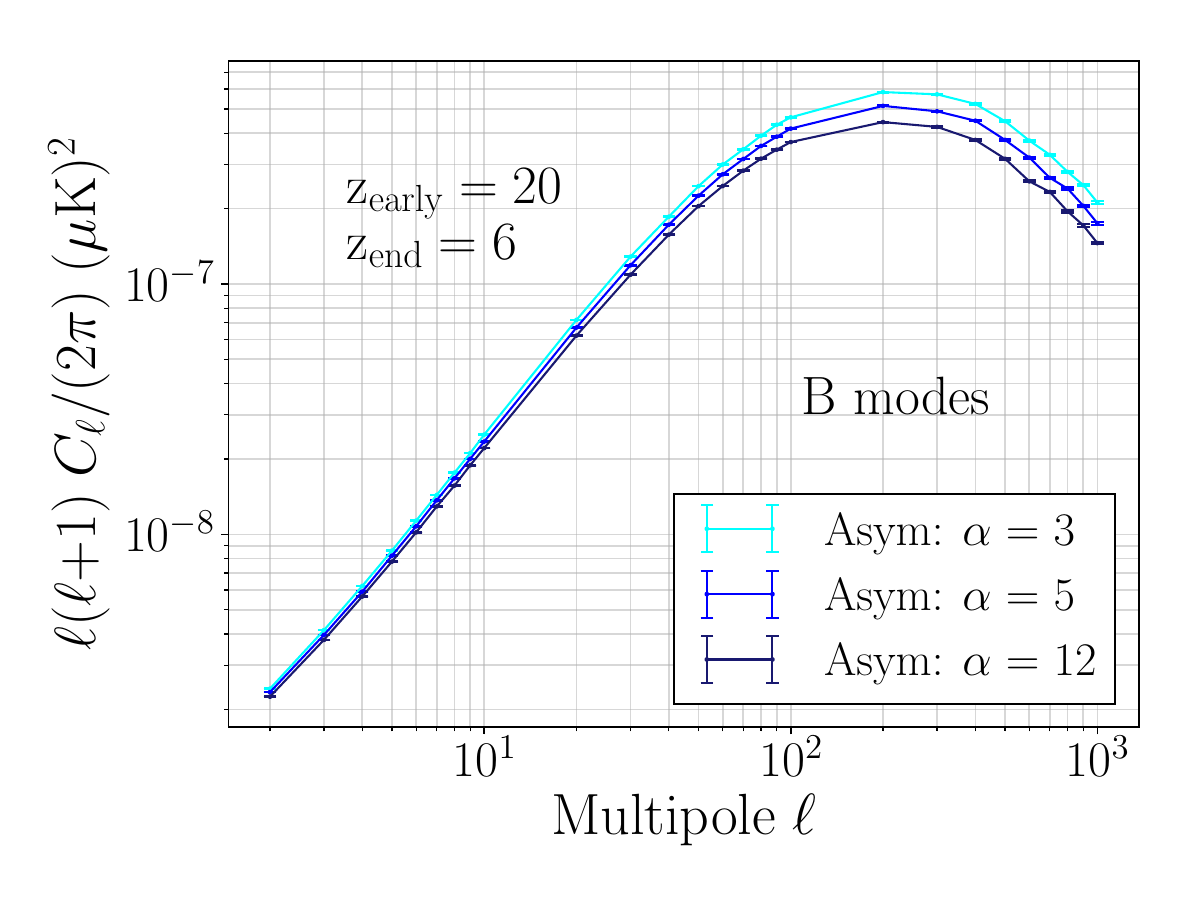}
		\caption{B modes}
		\label{fig:bmodes_asym}
	\end{subfigure}%
	\hspace{0.1cm}
	\begin{subfigure}{0.5\textwidth}
		\centering
		\includegraphics[width=1.08\linewidth]{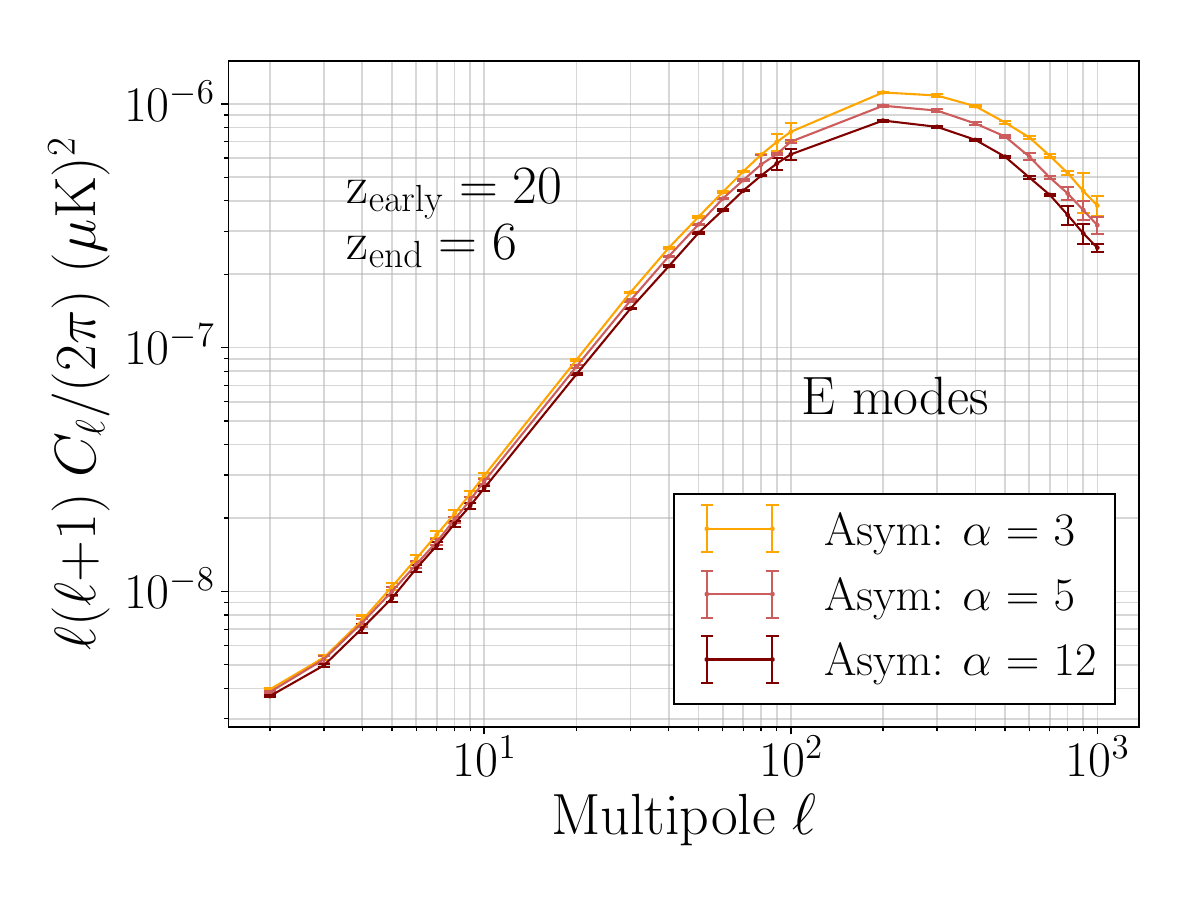}
		\caption{E modes}
		\label{fig:emode_asym}
	\end{subfigure}
	\vfill
		\hspace{-0.5cm}
	\begin{subfigure}{0.5\textwidth}
		\centering
		\includegraphics[width=1.08\linewidth]{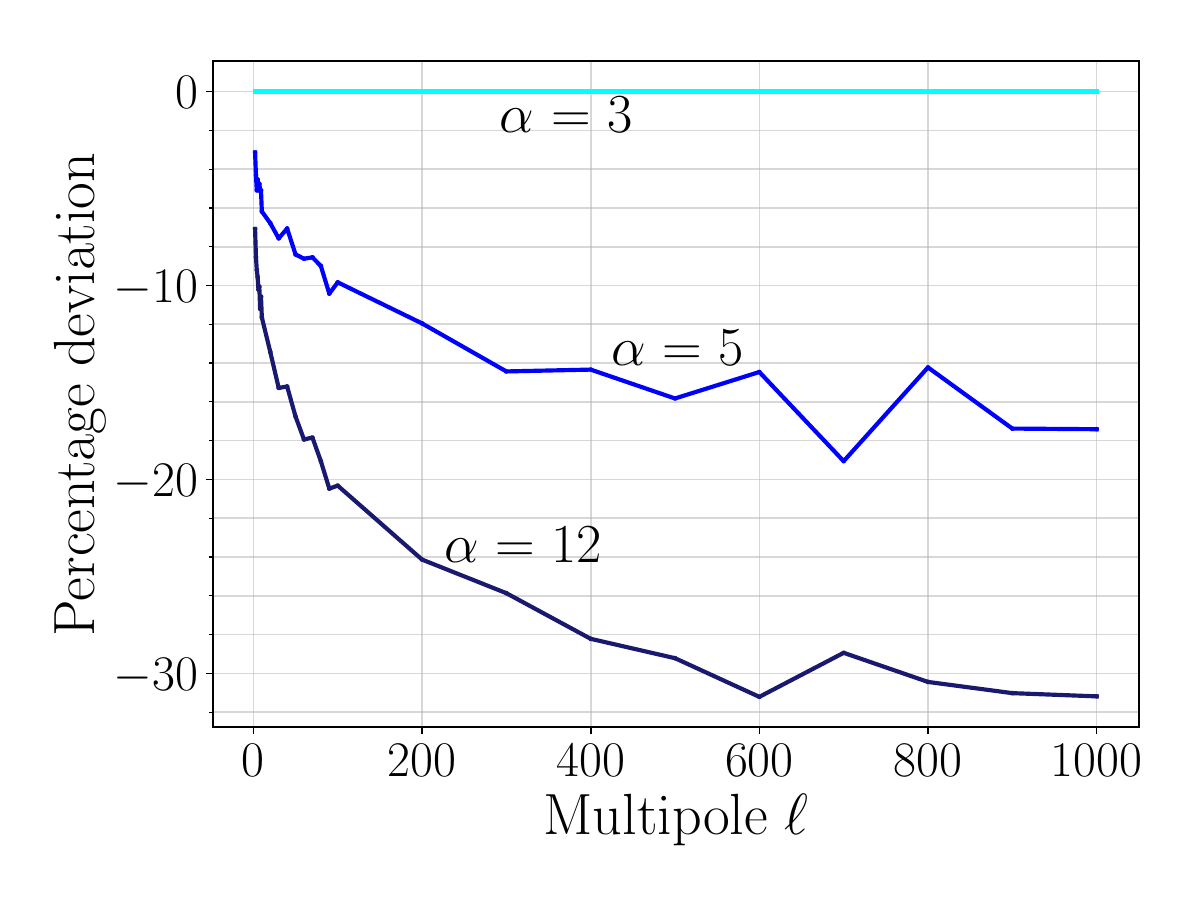}
		\caption{Percentage change in B modes}
		\label{fig:bmode_frac_asym}
	\end{subfigure}	
	\hspace{0.1cm}
	\begin{subfigure}{0.5\textwidth}
		\centering
		\includegraphics[width=1.08\linewidth]{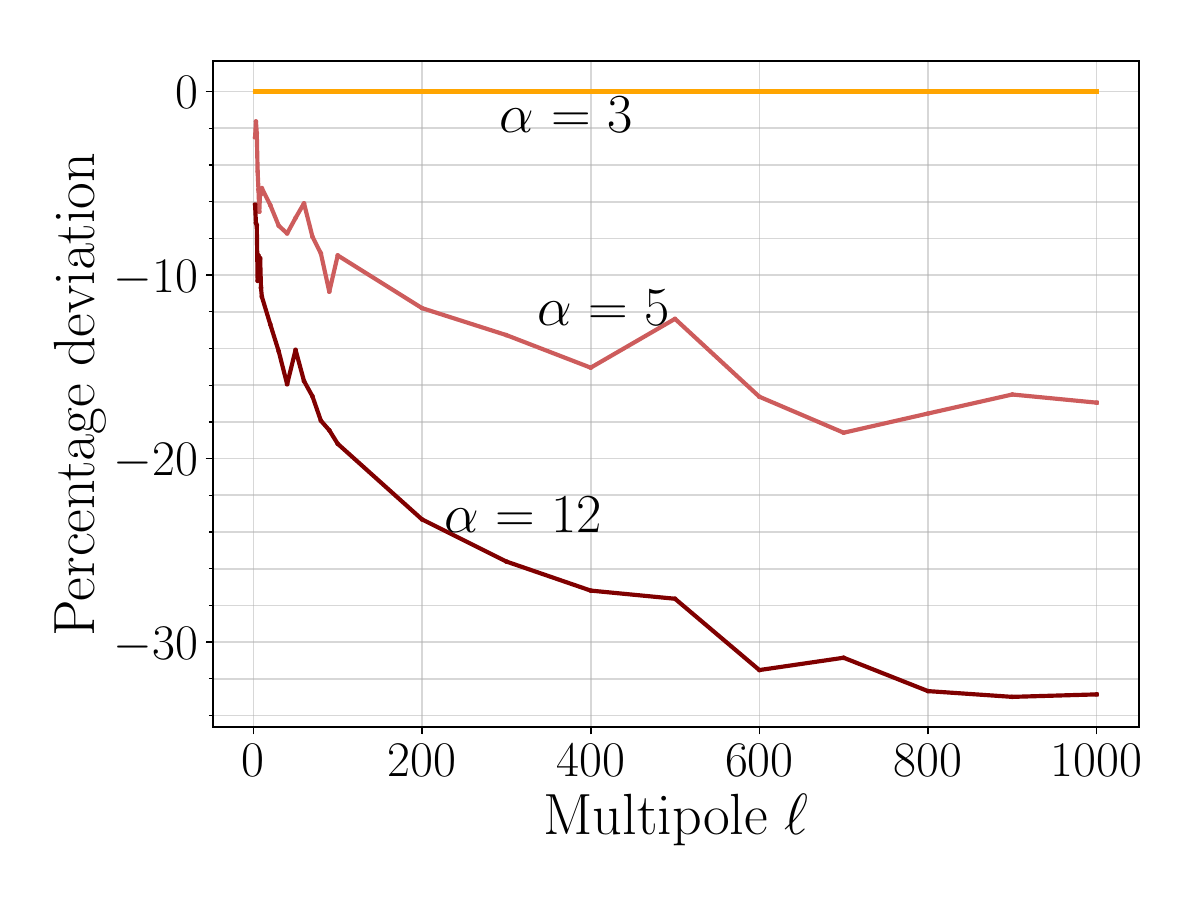}
		\caption{Percentage change in E modes}
		\label{fig:emode_frac_asym}
	\end{subfigure}	
	\caption{The angular power spectra for the case of asymmetric reionisation with $\alpha$ equal to 3, 5, and 12.  The optical depth increases with decrease in $\alpha$, boosting the power spectrum.}
	\label{fig:effect_redshift_asym}
\end{figure}With the increase in optical depth, as we see in figure \ref{fig:effect_redshift}, the polarisation signal also increases as expected. More scatterings between electrons and photons generate more polarisation, increasing the power spectrum at all scales smaller than the horizon size at reionisation. Changing the central redshift from 8.5 to 9.5 increases the optical depth by 17\% and it decreases by the same amount when the central redshift  is changed from 8.5 to 7.5. This get reflected in the power spectrum of the E and B modes. At the peak position, the power changes by the same amount as the optical depth, as seen in figure \ref{fig:bmode_frac_sym} and figure \ref{fig:emode_frac_sym}. In the case of asymmetric reionisation, with an increase in $\alpha$, reionisation happens later and faster as shown in figure \ref{fig:asym}. The total optical depth and the power spectra decrease as $\alpha$ increases. When $\alpha$ is changed from 3 to 12 the optical depth decreases by 34\%. We see the same percentage decrease in the power spectrum at the peak position in figure \ref{fig:bmode_frac_asym} and figure \ref{fig:emode_frac_asym}.
	
	
	\subsection{Dependence on duration of reionisation  \label{subsec_duration}}
\begin{figure}
			\hspace{-0.5cm}
		\begin{subfigure}{0.5\textwidth}
			\centering
			\includegraphics[width=1.08\linewidth]{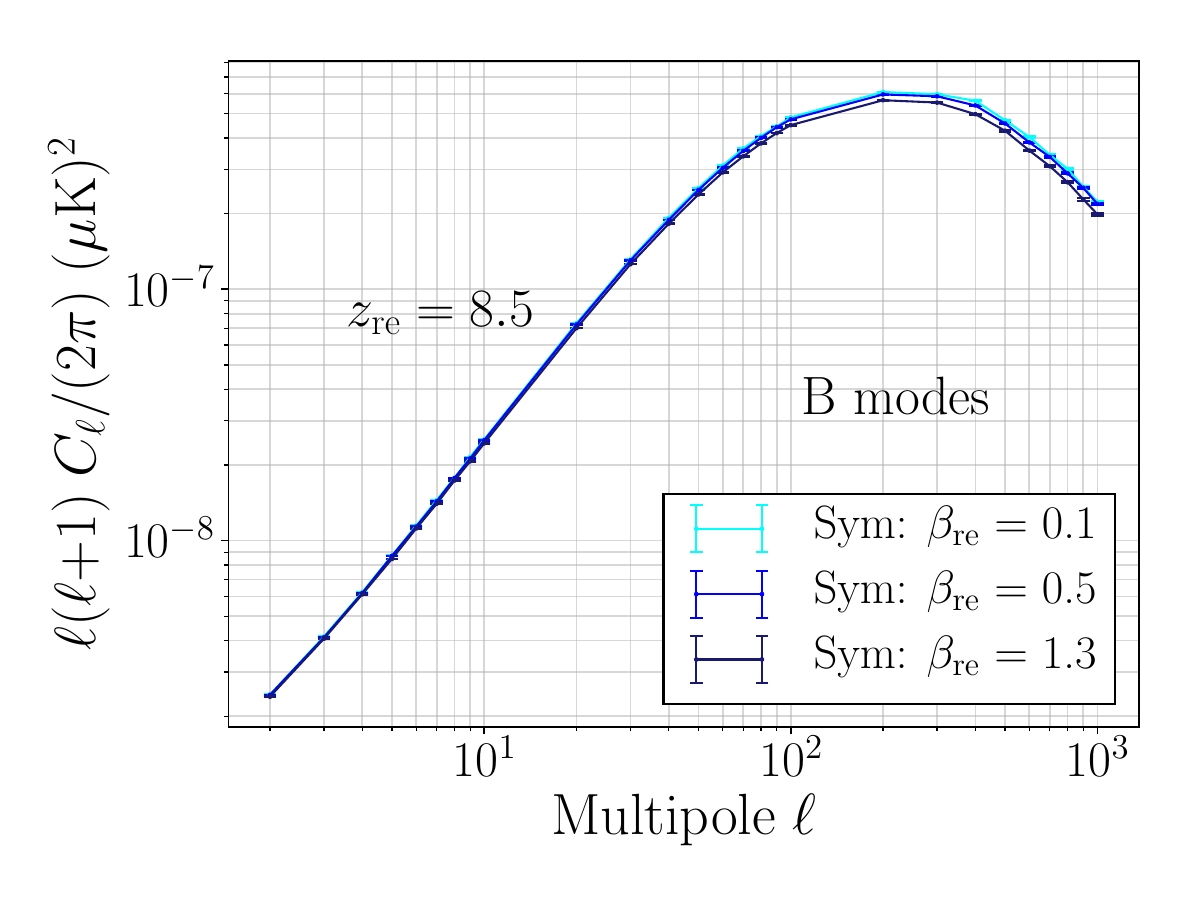}
			\caption{B modes}
			\label{fig:Bmode_width}
		\end{subfigure}%
		\hspace{0.1cm}
		\begin{subfigure}{0.5\textwidth}
			\centering
			\includegraphics[width=1.08\linewidth]{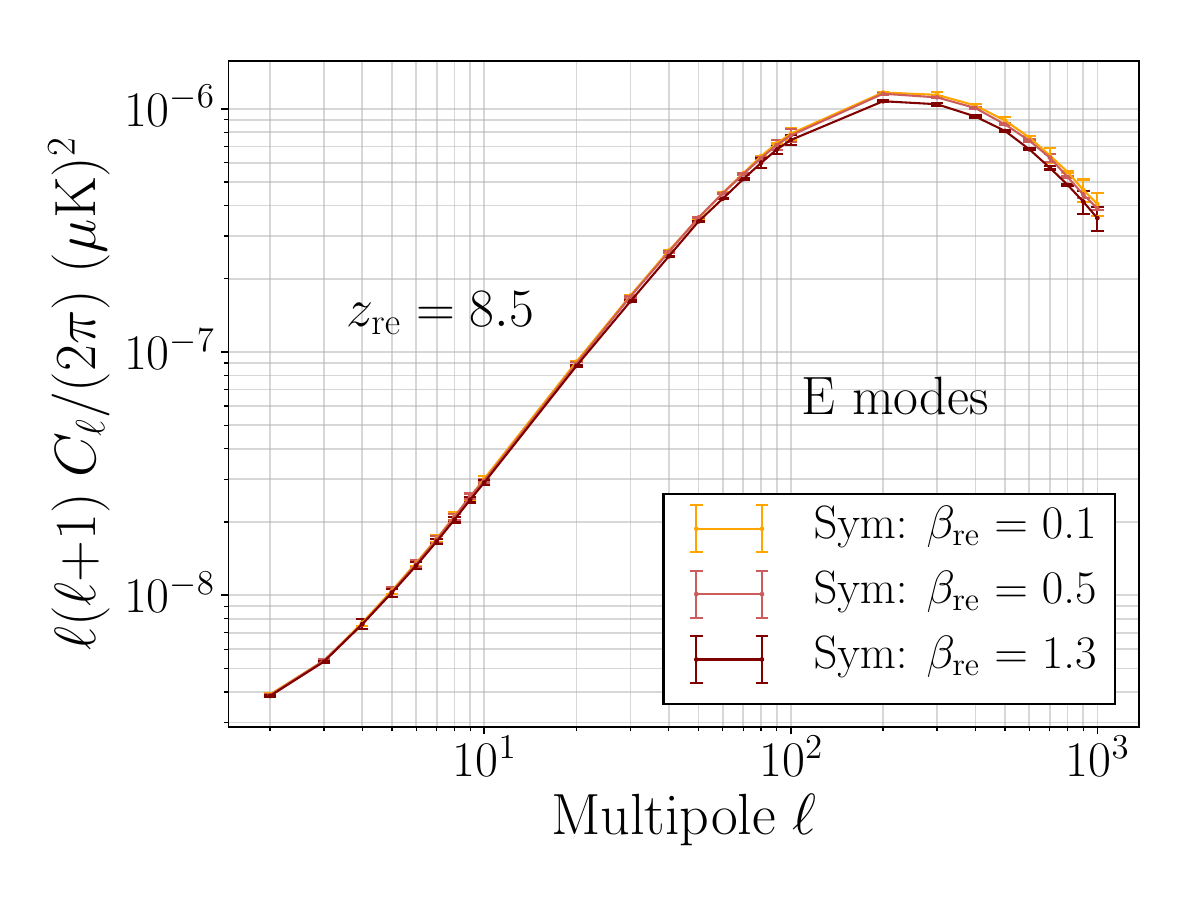}
			\caption{E modes}
			\label{fig:Emode_width}
		\end{subfigure}	
		\vfill
			\hspace{-0.55cm}
		\begin{subfigure}{0.5\textwidth}
			\centering
			\includegraphics[width=1.08\linewidth]{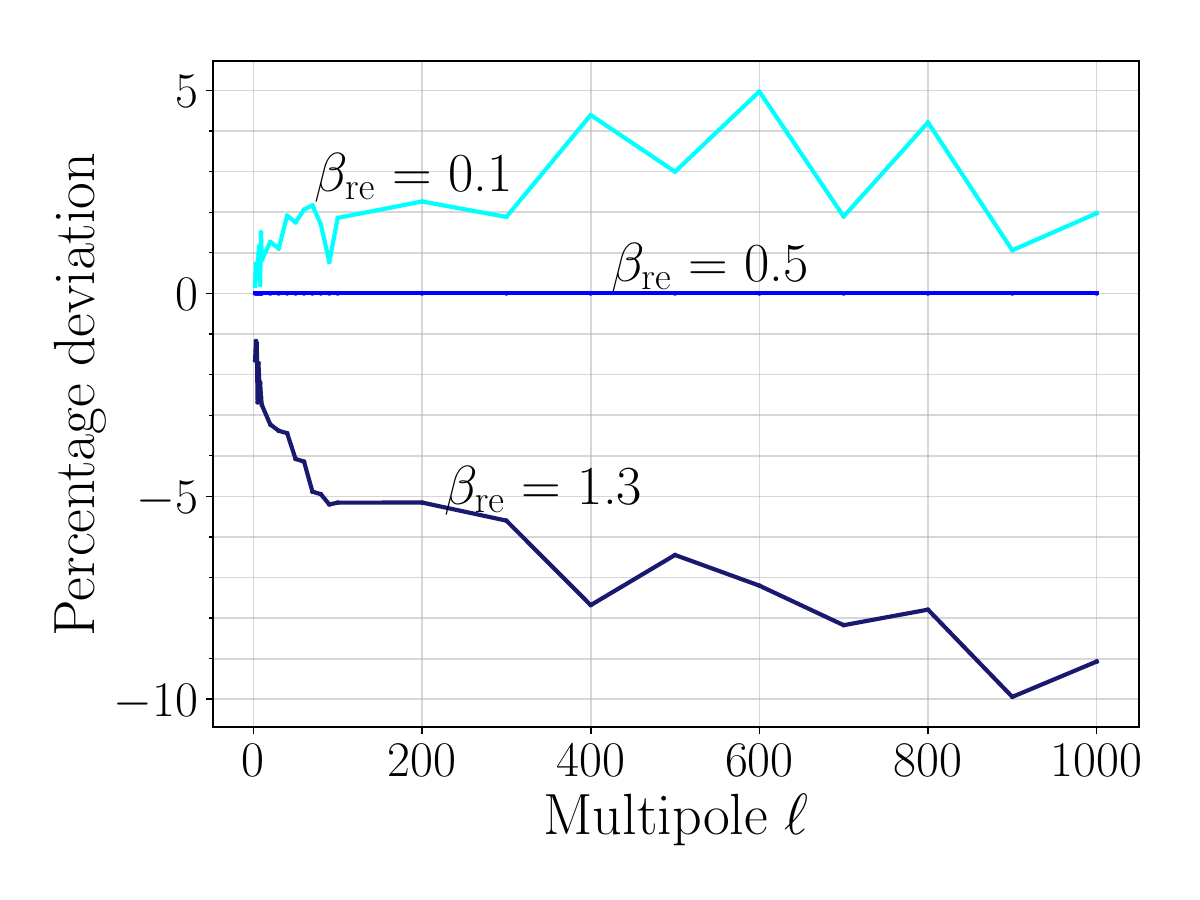}
			\caption{Percentage change in B modes}
			\label{fig:bmodesfrac}
		\end{subfigure}	
		\hspace{0.1cm}
		\begin{subfigure}{0.5\textwidth}
			\centering
			\includegraphics[width=1.08\linewidth]{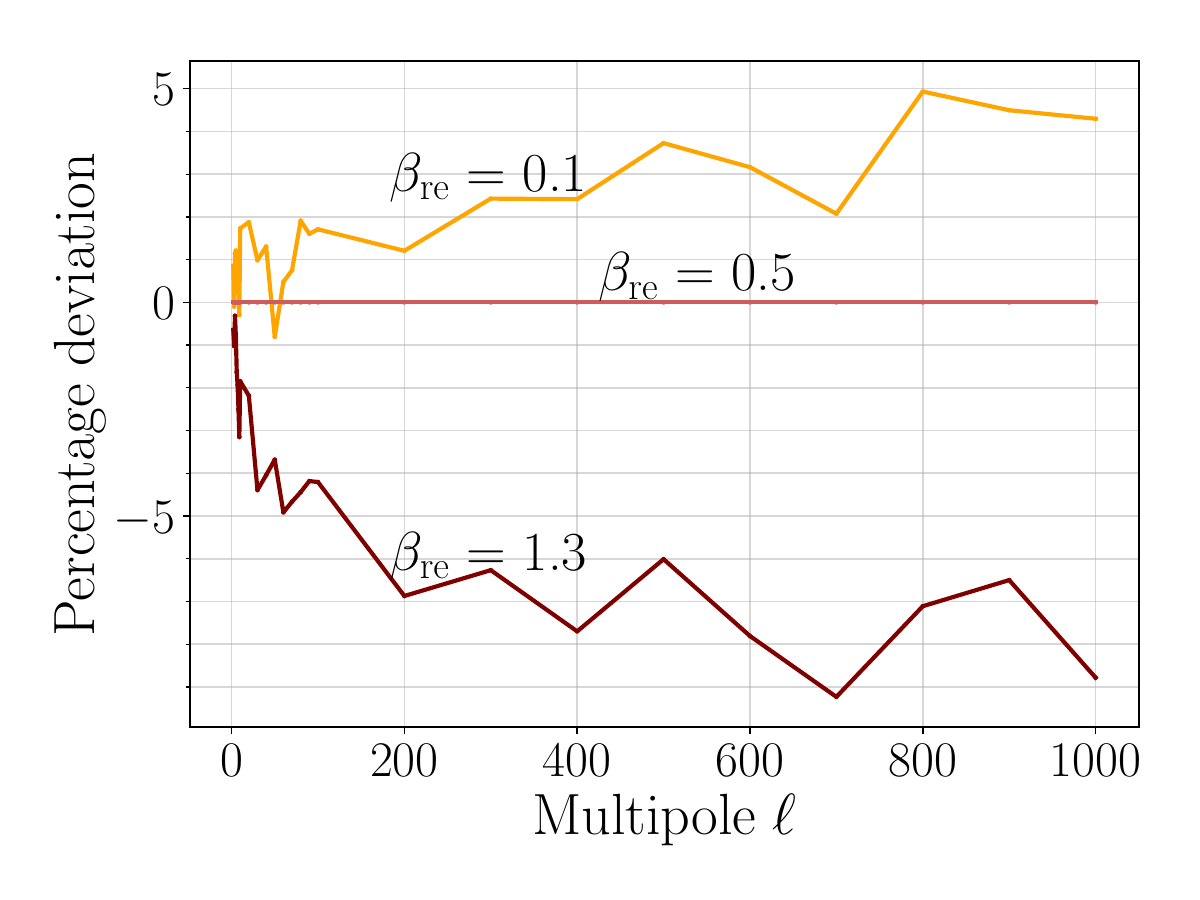}
			\caption{Percentage change in E modes}
			\label{fig:emodesfrac}
		\end{subfigure}	
		\caption{ The upper row shows the power spectra for different values of $\beta_{\mathrm{re}}$,  $\beta_{\mathrm{re}}=$  0.1 (almost instantaneous reionisation), 0.5, and 1.3 (gradual reionisation). The bottom panel shows the percentage change in power with respect to the case of reionisation with $\beta_{\mathrm{re}}=0.5$. }
		\label{fig:effect_width}
	\end{figure}

	When we keep the central redshift fix but change the duration of reionisation, the total optical depth changes negligibly. However, the power spectrum still changes by a significant amount. As shown in figure \ref{fig:Bmode_width} and figure \ref{fig:Emode_width}, the y-type angular power spectrum decreases with the increase in $\beta_{\mathrm{re}}$. 
	When the optical depth remains almost constant, the result of increasing the duration of reionisation is an increase in cancellation of polarisation created by uncorrelated velocity fields along the line of sight. This cancellation is greater on small scales as the number of uncorrelated regions within a given length is larger for modes with a smaller coherence scale. We thus expect that modes with scales smaller than the width of reionisation will be most significantly affected, with the suppression increasing with decreasing scale.  For example, consider $z_{\mathrm{re}}=8.5$, $\beta_{\mathrm{re}}=0.5$, and $\Delta z_{\mathrm{re}}=1.7283$, the comoving distance corresponding to the duration is, 
	\bea
	\label{duration}
	\Delta \chi=\chi\left(z=z_{10\%}\right)-\chi \left(z=z_{99\%}\right)=\chi\left(z=9.041\right)-\chi \left(z=7.313\right)\simeq 495 \mathrm{Mpc}.
	\eea
	This corresponds to a wave number of $\sim0.012 \mathrm{{Mpc}^{-1}}$, which in-turn corresponds to multipole of $\ell\sim 80$. Therefore, we expect power at $\ell \gtrsim 80$ to be suppressed compared to the case of almost instantaneous reionisation corresponding to $\beta_{\mathrm{re}}=0.1$. As the duration increases, this effect should become important at lower multipoles.  Indeed, this is what we observe from figure \ref{fig:bmodesfrac} and figure \ref{fig:emodesfrac}. 
	
	The pkSZ effect from reionisation was previously studied by
        Renaux-Petel et al. \cite{renaux2014spectral}. They reported not
        observing any noticeable difference in the power spectrum when they
        changed the duration of reionisation. They observed at
          most a 2\% difference at the peak of the spectrum when they
          changed the value of $\Delta q_{\mathrm{re}}$ from 0 to 3. They
          used the Planck 2013 parameters with best fit reionisation
          optical depth of 0.925 and central reionisation redshift of
          11.4. Changing the value of $\Delta q_{\mathrm{re}}$ from 0 to 3
          with $z_{\mathrm{re}}=11.4$ amounts to changing
          $\beta_{\mathrm{re}}$ from 0 to 0.56. At higher redshifts, the
          same reionisation width, $\Delta z_{\rm re}$, in redshift
          corresponds to a much smaller comoving distance. Thus the power
          spectrum is less sensitive to the change in reionisation duration
          $\Delta z_{\rm re}$ for higher reionisation redshift $z_{\rm re}$. However, the current Planck data \cite{adam2016planck} allows a parameter space where  $\beta_{\mathrm{re}}$ can reach almost 1.3 for $z_{\mathrm{re}}=8.5$. We observe an almost 3-10 \% difference at $\ell>10$ when we change $\beta_{\mathrm{re}}$  from 0.1 to 1.3 ($\Delta q_{\mathrm{re}} \rightarrow$ 0 to 6) with $\sim8\%$ difference at the peak at $\ell\sim200$. Compared to the fiducial case of $\beta_{\mathrm{re}}=0.5$, the power spectra for $\beta_{\mathrm{re}}=0.1$ and $\beta_{\mathrm{re}}=1.3$ show a deviation of 5\% at small scales as shown in figure \ref{fig:bmodesfrac} and figure \ref{fig:emodesfrac}.
	
	It is easy to understand why the polarisation signal should be sensitive to the width of the reionisation, as explained above, by looking at the cancellation of the polarisation coming from randomly oriented velocities of electrons along the line of sight. In fact, this cancellation is similar to the cancellation of the linear kSZ effect due to random orientation of the line of sight component of the velocities \cite{1987Vishniac}. We should therefore have similar sensitivity to reionisation in the linear kSZ effect which is much simpler to calculate. We explicitly calculate the effect of changing reionisation duration on the kSZ signal for comparison. The kSZ signal is proportional to the line of sight velocity field. The angular power spectrum is much simpler and has been calculated in the past \cite{Hernandez_Monteagudo_2006, hernandez2009peculiar}. Considering the same reionisation history and ignoring the spatial fluctuations in the electron density, the kSZ signal depends on the velocity-velocity two-point correlations. The angular  power spectrum is given by \cite{1987Vishniac,hernandez2009peculiar}:
	\begin{align}
		C^{\mathrm{kSZ}}_{\ell}=&\frac{\pi}{2}\sigma^2_{\mathrm{T}}\int_{0}^{\chi_{i}}d\chi\;e^{-\tau(\chi)}\, a(\chi)\int_{0}^{\chi_{i}}d\chi'\;e^{-\tau(\chi')}\,a(\chi)n_\mathrm{e}(\chi)n_\mathrm{e}(\chi')\int dkk^2 \, P_{uu}(k)\, j'_{\ell}(k\chi)j'_{\ell}(k\chi')
	\end{align}
	where $j_{\ell}'(x)=\frac{dj_{\ell}(x)}{dx}$. 
	\begin{figure}
		\hspace{-0.7cm}
		\vspace{0.35cm}
		\begin{subfigure}{0.5\textwidth}
			\hspace{-0.3cm}
			\begin{minipage}[c]{1.0\textwidth}
			\centering
			\includegraphics[width=1.08\linewidth]{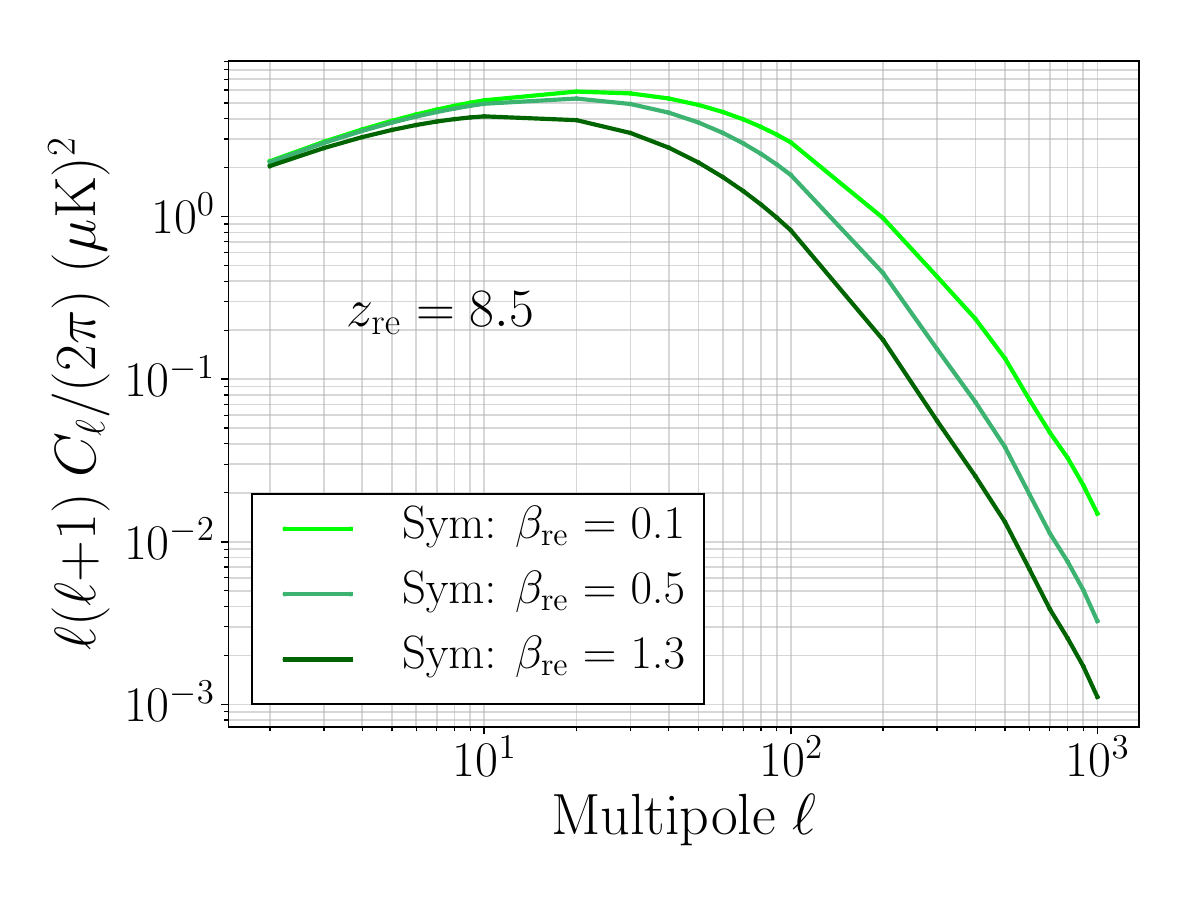}
			\end{minipage}
			\vspace{0.3cm}
			\caption{kSZ temperature angular power spectra.}
			\label{fig:polmodes_ksz}
		\end{subfigure}%
	\hspace{0.85cm}
		\begin{subfigure}{0.47\textwidth}
			\hspace{-0.95cm}
			\begin{minipage}[c]{1.07\textwidth}
			\centering
			\includegraphics[width=1.08\textwidth]{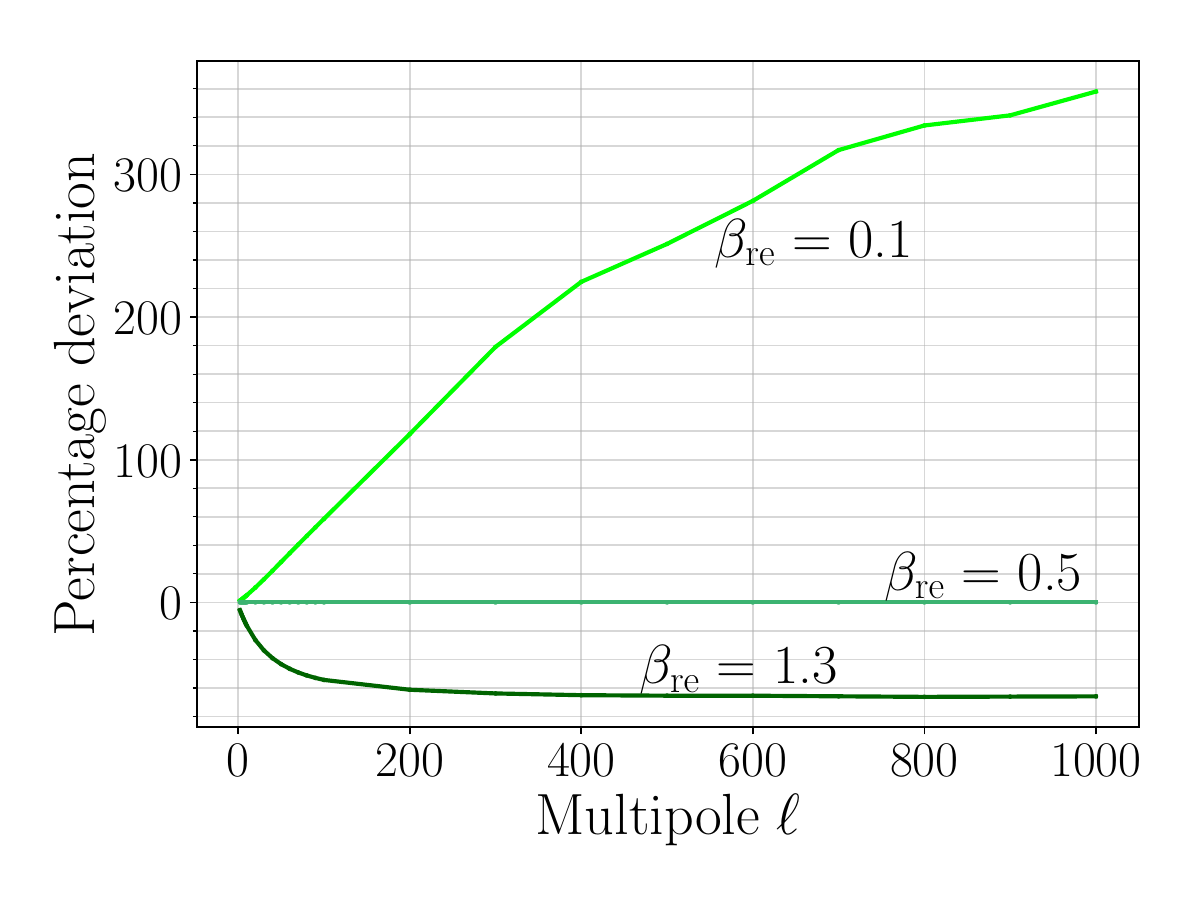}
			\end{minipage}
			\caption{Percentage change in power spectrum compared to the fiducial case of $\beta_{\mathrm{re}}=0.5$.}
			\label{fig:polemodes_frac_ksz}
		\end{subfigure}

		\caption{The left panel shows the kSZ temperature angular power spectrum for $ \beta_{\mathrm{re}}$ at 0.1, 0.5, and 1.3. We observe similar effects as in the case of polarisation. Though the optical depth change only by a small amount, the effect in the angular power spectrum is quite drastic. The power spectrum at small scales increases by 4 times for the case of almost instantaneous reionisation compared to our fiducial case of $ z_{\mathrm{re}}=8.5$ and $ \beta_{\mathrm{re}}=0.5$.}
		\label{fig:ksz_effects} 
	\end{figure}
	We have again considered the velocity field to be sourced by the scalar perturbations. We expect to see the same kind of suppression in the power spectrum at smaller scales when we increase the duration. Figure \ref{fig:ksz_effects} shows our results. Indeed we find very similar effects (in fact the effects are accentuated) on the power spectrum as seen in the polarisation signal. The power spectrum decreases by $\sim$ 80\% on small scales when we increase $\beta_{\mathrm{re}}$ from 0.1 to 1.3. Unfortunately, the linear kSZ signal is much weaker than the primary CMB signal and is overwhelmed by the cosmic variance of the primary CMB. So the linear kSZ power spectrum can never be observed. On the other hand, as mentioned before, the pkSZ effect has a different spectrum and is therefore not affected by the cosmic variance of primary CMB or the lensing E and B modes.
	\section{Contribution from Galaxy clusters shot noise \label{sec_shotnoise}}
	At low redshifts, $z\lesssim2$, most of the free electrons are in the intra cluster medium (ICM). We have so far calculated the contribution from the average number density of electrons and ignored the spatial fluctuations in the electron density. However, at $z\lesssim2$, there will be an extra contribution to the polarisation power spectrum due to the discrete nature of clusters of galaxies. Even if we assume that clusters are randomly distributed in space with some average number density, it should be noted that it is not a continuous density field. In any given small volume the number density of cluster and hence number density of free electrons will fluctuate following a Poisson distribution.  In the case of Poisson contribution, the spatial correlation decouples from the velocity correlation and we just modulate the expressions already derived for the polarisation power spectra. To model the cluster number density we consider a stochastic field $\Phi(\mathcal{M},\mathbf{r'},\eta)$ \cite{Hernandez_Monteagudo_2006, hernandez2009peculiar}, where $\mathcal{M}$ is the mass of the cluster, $\mathbf{r'}$ is the cluster center and $\eta$ is the conformal time. This field is Poisson distributed. Therefore, the correlation function becomes, 
	\begin{align}
	\Big\langle\Phi(\mathcal{M}_1,\mathbf{r_1'},\eta_1)\Phi(\mathcal{M}_2,\mathbf{r_2'},\eta_2)\Big\rangle=\bar{n}(\eta_1)\,\delta(\mathcal{M}_1-\mathcal{M}_2)\,\delta^{3}(\mathbf{r_1'}-\mathbf{r_2'}),
	\end{align}
	where $\bar{n}(\eta)$ is the mean number density at a conformal time $\eta$. It is given by the Sheth and Tormen mass function \cite{sheth2001ellipsoidal}. We also need to model the cluster gas profile by some window function $\mathrm{W}(\mathbf{r}-\mathbf{r'},\mathcal{M})$, where $\mathbf{r}$ is the comoving position vector. Since we are interested in the Poisson contributions on larger scales compared to cluster sizes, the exact profile is not important and we choose a Gaussian profile (see Appendix \ref{App:E_density}) for the window function. We can now write the electron number density as a function of both spatial and time coordinates,
	\bea
	\label{e_no_den_cluster}
	n_\mathrm{e}\left(\mathbf{ r},\eta \right) =\int d\mathbf{ r}{'}\int d\mathcal{M}\; n_\mathrm{e}^{0}(\mathcal{M}) \;\mathrm{W}\left(|\mathbf{ r}-\mathbf{ r}{'}|,\mathcal{M}\right)\Phi(\mathcal{M},\mathbf{ r}{'},\eta),
	\eea
	where $n_\mathrm{e}^{0}(\mathcal{M})$ is the central electron number density. Replacing the electron number density in eq.(\ref{pol_final}) using eq.(\ref{e_no_den_cluster}), we obtain the corresponding polarisation signal. Here, the density-density and velocity-velocity correlations are decoupled. After doing the same exercise as in the previous case we get the expressions for the Poisson contributions to the polarisation power spectra. They are given in Appendix \ref{App:Poisson}.
	\begin{figure}
		\hspace{-0.5cm}
	\begin{subfigure}{0.5\textwidth}
		\centering
		\includegraphics[width=1.08\linewidth]{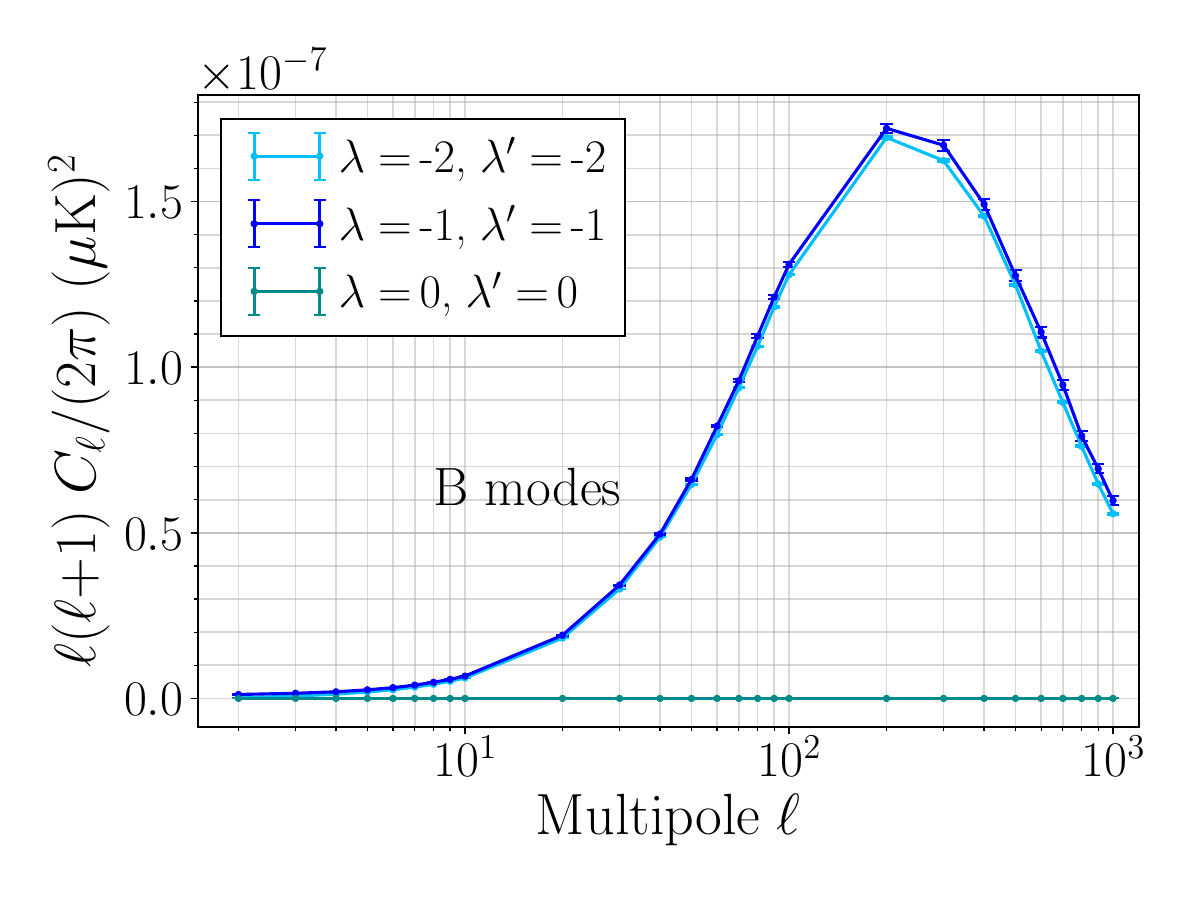}
		\caption{B modes: auto-correlation}
		\label{fig:Bmode_lambda_equal}
	\end{subfigure}%
	\hspace{0.1cm}
	\begin{subfigure}{0.5\textwidth}
		\centering
		\includegraphics[width=1.08\linewidth]{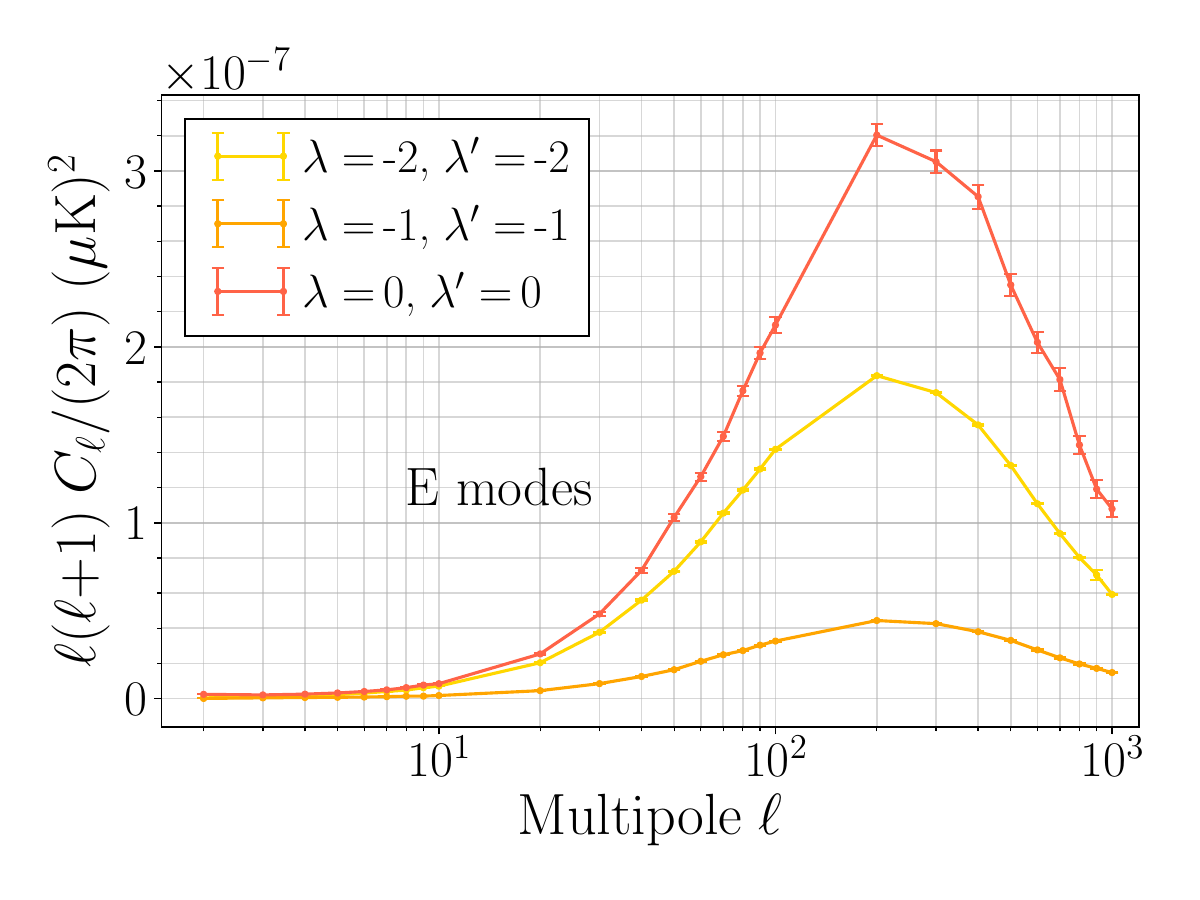}
		\caption{E modes: auto-correlation}
		\label{fig:Emode_lambda_equal}
	\end{subfigure}	
	\vfill
	\hspace{-0.5cm}
	\begin{subfigure}{0.5\textwidth}
		\centering
		\includegraphics[width=1.08\linewidth]{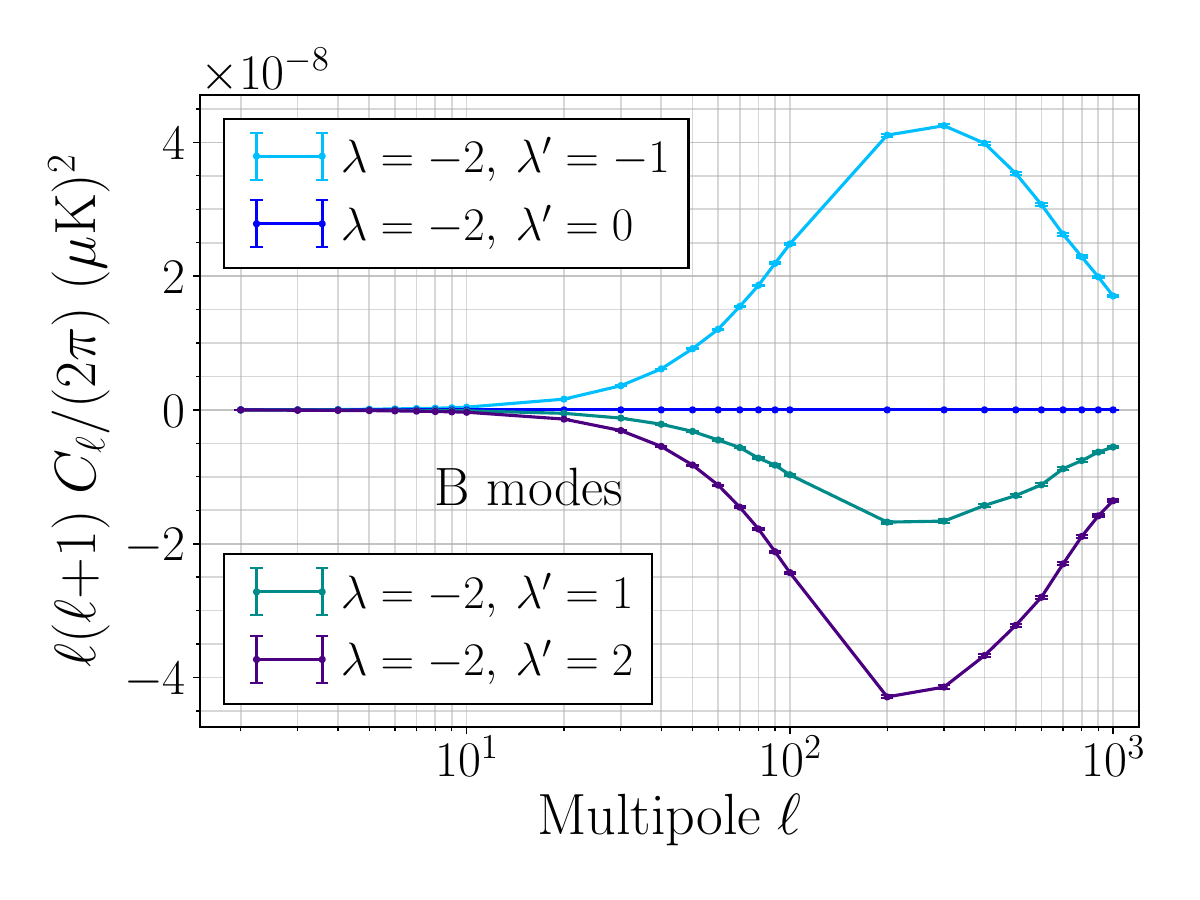}
		\caption{B modes: cross-correlation}
		\label{fig:Bmode_lambda_unequal}
	\end{subfigure}	
	\hspace{0.0cm}
	\begin{subfigure}{0.5\textwidth}
		\centering
		\includegraphics[width=1.08\linewidth]{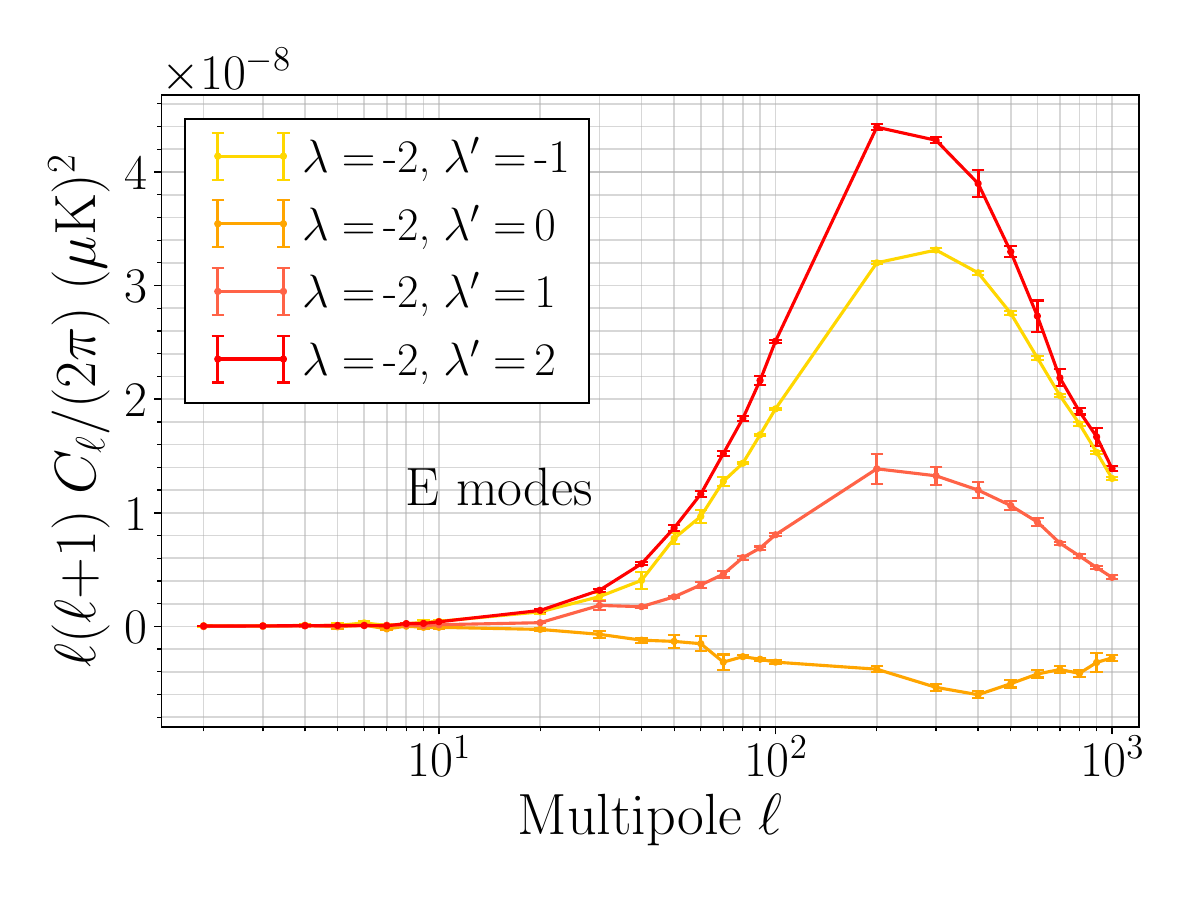}
		\caption{E modes: cross-correlation}
		\label{fig:Emode_lambda_unequal}
	\end{subfigure}	
	\caption{The upper panel shows the contribution from scalar, vector, and tensor auto correlations. The lower panel shows the same for cross-correlations. There is no B mode polarisation from scalar modes. For E modes the scalar modes contribution is the highest. This is the reason behind the greater power in E modes than in B modes. Also, the tensor mode $(\lambda$ and $\lambda'=\pm2)$ contribution is roughly equal for both the modes, while the vector mode $(\lambda$ or $\lambda'=\pm1)$ contributes dominantly to the B modes. .}
	\label{fig:Diff_lambda}
\end{figure}
	The contribution from these terms is shown in figure \ref{fig:summary plot}. We observe that the signal from the cluster shot noise is on average 2 orders of magnitude less than the contributions from reionisation on large scales but becomes comparable on small scales. We have only considered the Poisson contribution from the clusters. The clusters of galaxies are however also clustered spatially. We have also ignored the fact that reionisation is expected to be patchy as there can be large fluctuation in the electron density field during reionisation. Taking the electron number density fluctuations fully into account will give rise to terms which are formally $3^{\mathrm{rd}}$ order in perturbation theory and the power spectra will involve 6-point correlation functions. Our focus in this paper is on $2^{\mathrm{nd}}$ order terms and we leave the higher order terms for future work. We should however point out that even though patchy reionisation is formally a $3^{\mathrm{rd}}$ order effect, it does not mean that the contribution would be small. The electron density fluctuations can be of order unity and some of the cancellations in the $2^{\mathrm{nd}}$ order terms can be avoided, similar to patchy kSZ effect \cite{1987Vishniac}. Thus patchy reionisation may give a comparable contribution. Calculating the power spectra, however, requires dealing with the product of 3 perturbation variables instead of 2 and is more complicated and numerically challenging. We plan to study these effects in a separate paper in the near future. 
	
	\section{Scalar, Vector, and Tensor  contributions to the power spectrum \label{subsec_SVT_contribution} }
	From all the plots of the E and B modes, it is evident that the E modes are always greater than the B modes. A way to investigate why this is the case is to look at individual contributions from scalar, vector and tensor components of the polarisation field. It should be noted that although the velocity field is sourced by purely scalar field, at second order, scalar, vector and tensor components are all present. Different components corresponds to choosing different values of $\lambda$ in eq.(\ref{pol_final}). In this way, we can also check whether a cross correlation between different components producing a specific type of polarisation exists or not and if it does then to what extent? To see this we choose different values of $\lambda$ and $\lambda'$ in eq.(\ref{cl_ee_2}) and eq.(\ref{cl_bb_2}) and plot the corresponding power spectrum.  The results are shown in figure \ref{fig:Diff_lambda}. As expected, the B-modes from the scalar modes, i.e. $\lambda$ or $\lambda'=0$ vanish. In fact the integrand vanishes identically when  $\lambda$ or $\lambda'=0$ in the case of B modes.  We also observe that the cross correlations between scalar, vector and tensor modes are much smaller than the auto-correlations. For B modes, the vector and the tensor modes contribute equally whereas for E modes the maximum contribution is from the scalar modes and the least from the vector modes. Also, we see that the tensor modes have almost an equal contribution to both the E and B modes, while the vector modes primarily contribute to the B modes. We refer the reader to \cite{HU1997primer} for a more detailed discussion.

	\section{Conclusion \label{sec_conclusion}}
	We have calculated the y-type E and B mode polarisation angular power spectrum arising from the transverse velocity of free electrons during the reionisation and post-reionisation eras. We have shown that the polarisation signal is not only sensitive to the central redshift of reionisation or optical depth as expected but more interestingly to the duration of reionisation also. Our results, through a more detailed study, show that the conclusion drawn by Renaux-Petel et al. \cite{renaux2014spectral} about the dependence of polarisation power spectra on the duration of reionisation is not complete and it stems from the fact that the whole allowed parameter space was not explored. We point out the close relationship between the linear kSZ effect, which probes the line of sight component of the electron velocity and the pkSZ effect. The response of the pkSZ effect to the duration of reionisation is similar to the kSZ effect.
	 
	We want to mention that as we were completing this work another paper on the pkSZ effect appeared on arXiv \cite{pksZ_Kamion}.  They also studied the polarisation signal from reionisation, but their work was focused on using this signal to probe cosmic birefringence and non-Gaussianity rather than reionisation. Our numerical results for reionisation agree qualitatively and are of similar magnitude, although they do not specify the exact reionisation history they have used. Our expressions for the E and B mode power spectrum are equivalent, although written in a different form. We show the equivalence in  Appendix \ref{App:Quad_dep}. 
	
	We have not included the spatial fluctuations in the electron density field, in particular patchy reionisation, in our analysis. This will formally include $3^{\mathrm{rd}}$ order terms. We leave the higher order calculations for our future work. Similarly, for the contributions from galaxy clusters, we also need to include the effects of spatial clustering of galaxy clusters.  We however expect the contribution from galaxy clusters to be sub-dominant in analogy with the linear kSZ effect.
	
	The spectrum of the pkSZ effect can be decomposed into a sum of differential blackbody spectrum (identical to the primary CMB anisotropies) and a y-type spectrum. This is very important from the component separation perspective. We can extract the pkSZ effect from the multi-frequency CMB data by separating the y-type signal while suppressing the blackbody signal. This strategy can enable us to detect the pkSZ signal unencumbered by the cosmic variance of the primary CMB anisotropies. Moreover, the other dominant y-type distortion anisotropies, primary as well as secondary such as the thermal SZ effect, are unpolarised. Thus we can, in principle, measure this signal as precisely as the primary CMB signal if sufficient sensitivity is reached in future. The pkSZ effect has important cosmological information. It is sensitive to the matter velocity power spectrum in addition to the parameters of reionisation. The E and B modes of the y-type distortions thus have the potential to measure the cosmological parameters beyond the cosmic variance limit of the blackbody CMB anisotropies. 
	\acknowledgments
	This work is supported by the Department of Atomic Energy, Government of India, under Project Identification Number RTI 4002.
	This work is also supported by Max Planck Partner Group for the cosmology of Max Planck Institute for Astrophysics Garching at Tata Institute of Fundamental Research funded by Max-Planck-Gesellschaft. We acknowledge the use of computational facilities of the Department of Theoretical Physics at Tata Institute of Fundamental Research, Mumbai. AKG is thankful to Carlos Hernández-Monteagudo, Aseem Paranjape, Subhabrata Majumdar, and Anoma Ganguly for useful discussions.
	\appendix
	\section{Reionisation models \label{App:Reion_history}}
\begin{figure}
	\vspace{-1.0cm}
	\hspace{-0.6cm}
	\begin{subfigure}{0.59\textwidth}
		\centering
		\includegraphics[width=1.07\linewidth]{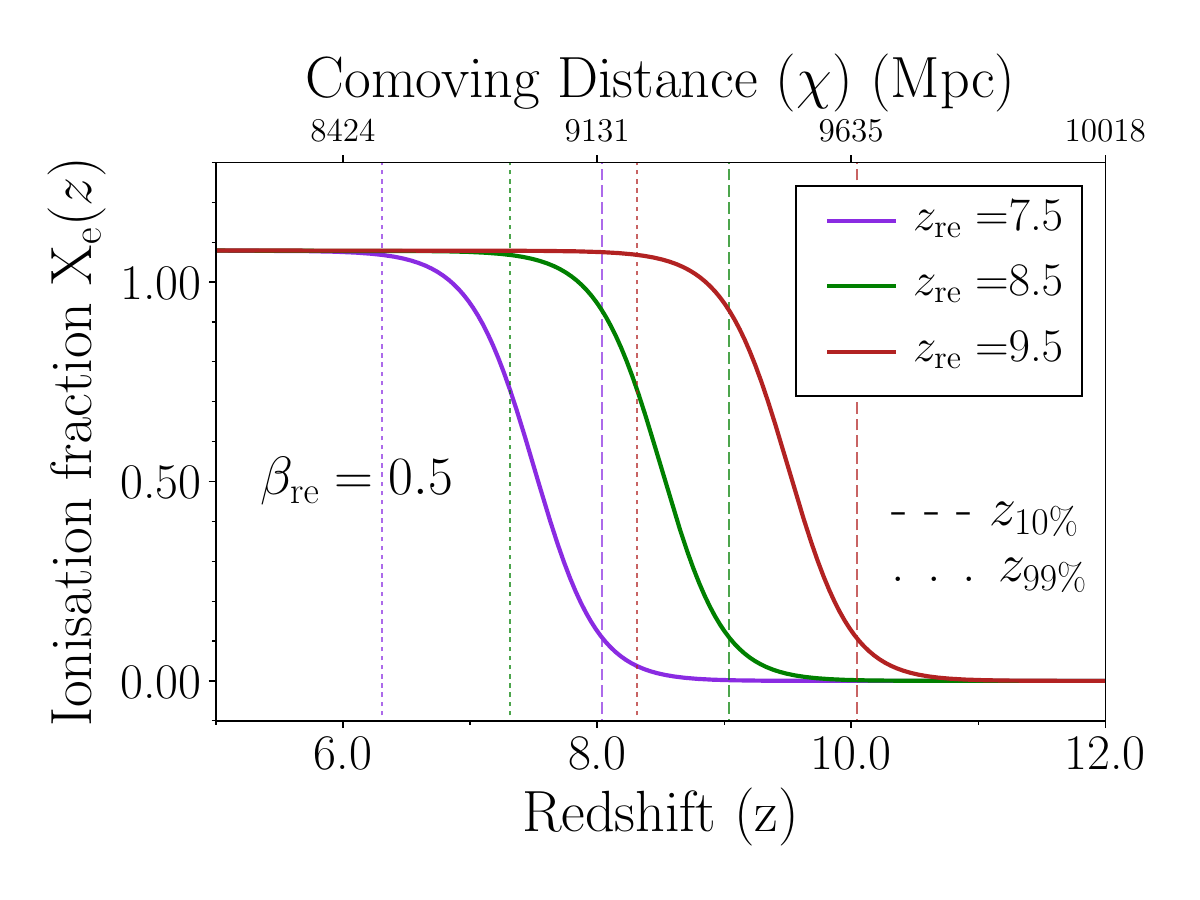}
		\vspace{-0.92cm}
		\caption{At a fixed duration. }
		\label{fig:sym_centralz}
	\end{subfigure}%
		\hspace{0.59cm}
			\begin{subtable}{0.85\textwidth}
				\vspace{0.0cm}
				\hspace{-0.11cm}
				\begin{minipage}[c]{0.43\textwidth}
				
					\begin{tabular}{|ccc|}
						\hline
						\multicolumn{3}{|c|}{At $\beta_{\mathrm{re}}= 0.5$}                                                                                                                                                                                                 \\ \hline
						\multicolumn{1}{|c|}{\begin{tabular}[c]{@{}c@{}}Central \\Redshift $(z_{\mathrm{re}})$\end{tabular}} & \multicolumn{1}{c|}{\begin{tabular}[c]{@{}c@{}} $z_{10\%}$\end{tabular}} & $z_{99\%}$ \\ \hline
						\multicolumn{1}{|c|}{7.5}                                                                                  & \multicolumn{1}{c|}{8.040}                                                                                     & 6.308             \\ \hline
						\multicolumn{1}{|c|}{8.5}                                                                                  & \multicolumn{1}{c|}{9.041}                                                                                     & 7.313             \\ \hline
						\multicolumn{1}{|c|}{9.5}                                                                                  & \multicolumn{1}{c|}{10.042}                                                                                     & 8.317            \\ \hline
					\end{tabular}
					\caption{For different central redshift, fixing $\beta_{\mathrm{re}}= 0.5$.}
					\label{tab: table4 }
				\end{minipage}
			\end{subtable}
		
		\hspace{-0.6cm}
	\begin{subfigure}{0.59\textwidth}
		\centering
		\includegraphics[width=1.07\linewidth]{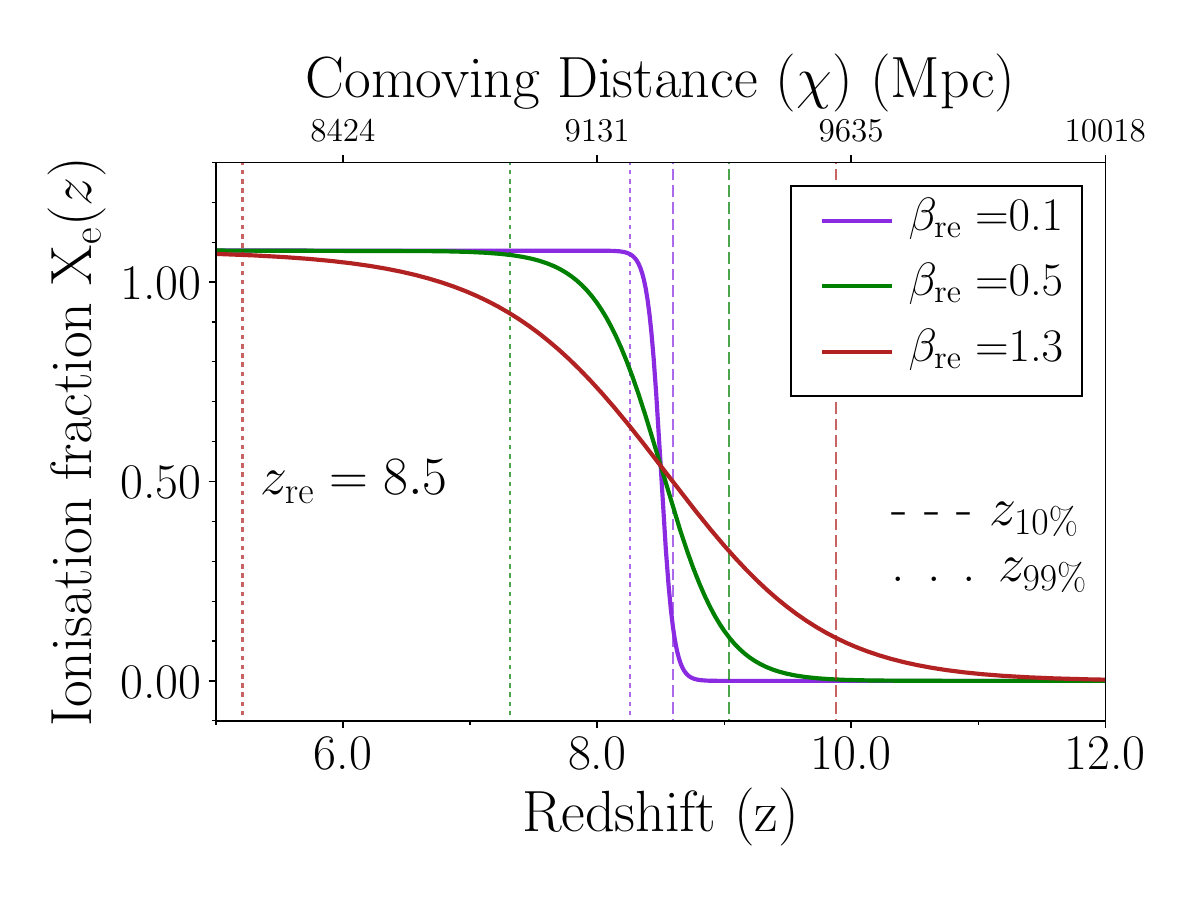}
			\vspace{-0.92cm}
		\caption{At a fixed central redshift.}
		\label{fig:sym_duration}
	\end{subfigure}	%
		\hspace{0.79cm}
		\begin{subtable}{1.5\textwidth}
			\vspace{0.0cm}
			\hspace{-0.41cm}
		\begin{minipage}[c]{0.22\textwidth}

			\begin{tabular}{|ccc|}
				\hline
				\multicolumn{3}{|c|}{At central redshift $z_{\mathrm{re}}=8.5$}                                                                                                               \\ \hline
				\multicolumn{1}{|c|}{$\beta_{\mathrm{re}}$} & \multicolumn{1}{c|}{\begin{tabular}[c]{@{}c@{}} $z_{10\%}$\end{tabular}} & $z_{99\%}$ \\ \hline
				\multicolumn{1}{|c|}{0.1}                   & \multicolumn{1}{c|}{8.60}                                                                               & 8.262           \\ \hline
				\multicolumn{1}{|c|}{0.5}                   & \multicolumn{1}{c|}{9.041}                                                                               & 7.313          \\ \hline
				\multicolumn{1}{|c|}{1.3}                   & \multicolumn{1}{c|}{9.879}                                                                               & 5.209            \\ \hline
			\end{tabular}
			\caption{For different width of reionisation, fixing central redshift at $z_{\mathrm{re}}=8.5$.}
			\label{tab: table5 }
		\end{minipage}
	\end{subtable}

	\hspace{-0.6cm}
	\begin{subfigure}{0.59\textwidth}
		\centering
		\includegraphics[width=1.07\linewidth]{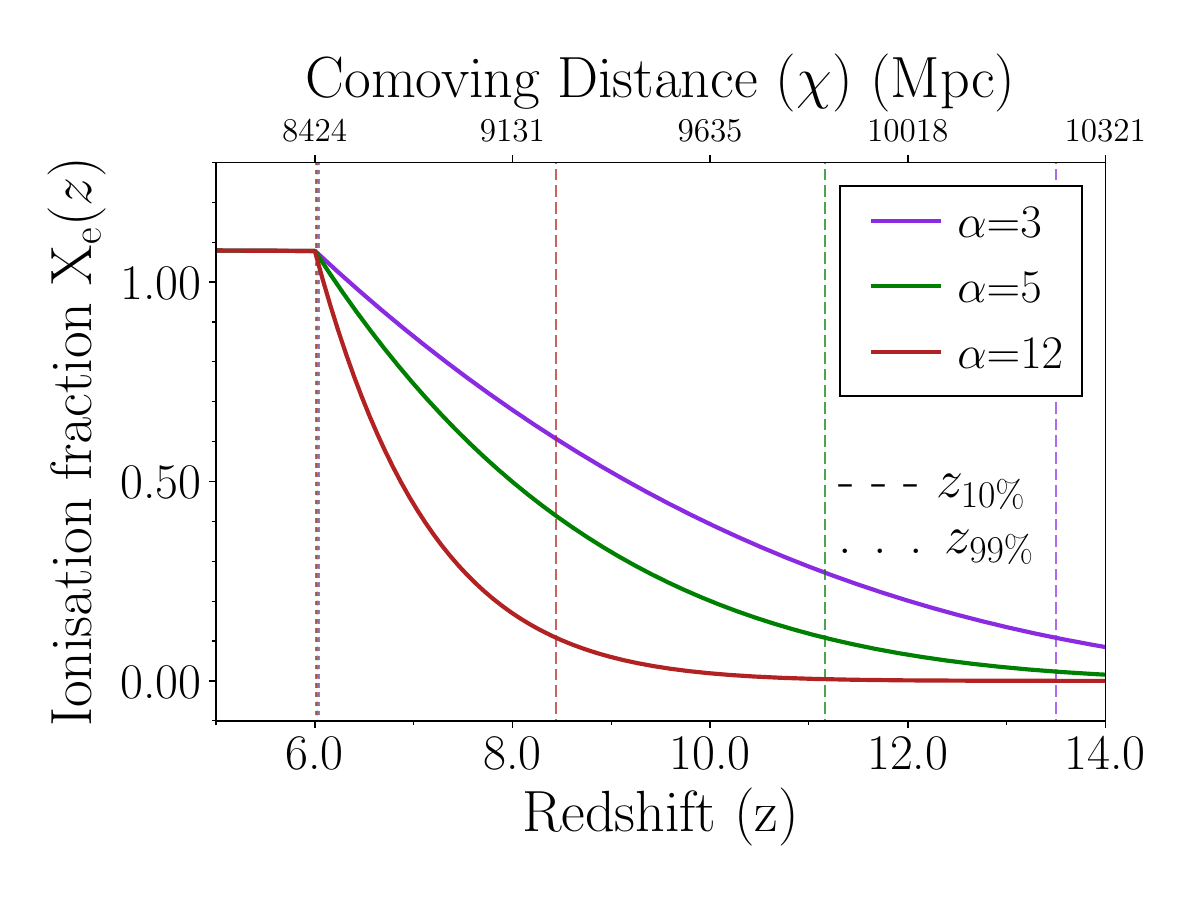}
			\vspace{-0.92cm}
		\caption{For different rapidity parameter $\alpha$.}
		\label{fig:asym}
	\end{subfigure}	%
	\hspace{0.79cm}
		\begin{subtable}{1.5\textwidth}
			\vspace{0.0cm}
			\hspace{-0.41cm}
			\begin{minipage}[c]{0.22\textwidth}
		\begin{tabular}{|ccc|}
			\hline
			\multicolumn{3}{|c|}{At $\mathrm{z_{early}}=20$ and $\mathrm{z_{end}}=6$}                                                                                        \\ \hline
			\multicolumn{1}{|c|}{$\alpha$} & \multicolumn{1}{c|}{\begin{tabular}[c]{@{}c@{}}$z_{10\%}$\end{tabular}} & $z_{99\%}$ \\ \hline
			\multicolumn{1}{|c|}{3}        & \multicolumn{1}{c|}{13.50}                                                                               & 6.046            \\ \hline
			\multicolumn{1}{|c|}{5}        & \multicolumn{1}{c|}{11.16}                                                                               & 6.028           \\ \hline
			\multicolumn{1}{|c|}{12}       & \multicolumn{1}{c|}{8.44}                                                                               &6.011             \\ \hline
		\end{tabular}
		\caption{For different rapidity parameter $\alpha$, fixing $\mathrm{z_{early}}$ = 20 and $\mathrm{z_{end}}$ = 6.}
		\label{tab: table6 }
	\end{minipage}
\end{subtable}
	\caption{The ionisation fraction $\mathrm{X_e}(z)$ is shown for different parameters that have been used to plot the power spectra. On the top-left, we have shown $\mathrm{X_e}(z)$ for different central redshift reionisation, while on the center-left $\mathrm{X_e}(z)$ for different duration of reionisation has been shown. The bottom plot is for asymmetric reionisation for different rapidity parameter $\alpha$.}
	\label{fig:reion_history}
\end{figure}

We show in figure \ref{fig:reion_history}  the ionisation fraction, $\mathrm{X_e}(z)$, for the different sets of reionisation parameters that were used in our analysis. Figure \ref{fig:sym_centralz} and figure \ref{fig:sym_duration} show the ionisation fraction for the case of symmetric reionisation, for different central redshift $z_{\mathrm{re}}$ keeping $\beta_{\mathrm{re}}$ fixed and vice-versa. Also shown are the redshifts corresponding to $z_{10\%}$ and $z_{99\%}$. For the case of a fixed $\Delta z_{\mathrm{re}}$, though the redshift interval of the duration is fixed, it corresponds to a different physical time interval as an equal redshift interval centred around earlier epochs correspond to shorter physical time interval. The case of asymmetric reionisation is shown in figure \ref{fig:asym} for different rapidity parameters $\alpha$. 
\section{Derivation of harmonic coefficients and angular power spectrum \label{App:EB_coeff_power}}
	This section contains all the steps to derive the harmonic coefficients. We begin with eq.(\ref{harm_coeff_start}).
	\begin{align}
		a_{\ell m}=&4\pi\left(\frac{4\pi}{3}\right)^2\sqrt{\frac{3}{2\pi}}\frac{\sqrt{6}\sigma_{\mathrm{T}}}{10}\sum_{\lambda =-2}^{2}(-1)^\lambda \int_{0}^{\chi_{i}}d\chi\;e^{-\tau(\chi)}\,n_\mathrm{e}(\chi)\,a(\chi)\int \int \frac{d^3\mathbf{k_1}d^3\mathbf{k_2}}{(2\pi)^6} \tilde{u}(\mathbf{k_{1}})\tilde{u}(\mathbf{k_{2}})\nonumber\\
		& \sum_{{L, M}}i^L Y_{L M}^{*}(\hat{\mathbf{k}})\,j_{L}(k\chi)\sum_{p_1,p_2}	\left(\begin{array}{ccc}
			1& 1 & 2\\ 
			p_1& p_2 & -\lambda
		\end{array}\right)Y_{1p_1}^{*}(\mathbf{{\hat{k}}_1})Y_{1p_2}^{*}(\mathbf{{\hat{k}_2}})\;A^{\lambda L M}_{\ell m}.
	\end{align}
	To find the E and B modes coefficient we need to know $a^{*}_{\ell m}$. Taking the complex conjugate of $a_{\ell m}$ we get
	\begin{align}
		a^{*}_{\ell m}=&4\pi\left(\frac{4\pi}{3}\right)^2\sqrt{\frac{3}{2\pi}}\frac{\sqrt{6}\sigma_{\mathrm{T}}}{10}\sum_{\lambda =-2}^{2}(-1)^\lambda \int_{0}^{\chi_{i}}d\chi\;e^{-\tau(\chi)}\,n_\mathrm{e}(\chi)\,a(\chi)\int \int \frac{d^3\mathbf{k_1}d^3\mathbf{k_2}}{(2\pi)^6} \tilde{u}^{*}(\mathbf{k_{1}})\tilde{u}^{*}(\mathbf{k_{2}})\nonumber\\
		&\sum_{{L, M}}(-i)^L Y_{L M}(\hat{\mathbf{k}})\,j_{L}(k\chi)\sum_{p_1,p_2}
		\left(\begin{array}{ccc}
			1& 1 & 2\\ 
			p_1& p_2 & -\lambda
		\end{array}\right)Y_{1p_1}(\mathbf{{\hat{k}}_1})Y_{1p_2}(\mathbf{{\hat{k}_2}})\;A^{\lambda L M}_{\ell m}.
	\end{align}
	Since the velocity potentials are Gaussian random fields, $\tilde{u}^{*}(\mathbf{k})=\tilde{u}(\mathbf{-k})$. Now, we can change all the $\mathbf{k}$ vectors to $-\mathbf{k}$ vectors. Since  the integral is over all the $\mathbf{k}$ space, the limits of integration do not change. Doing so we get
	\begin{align}
		a^{*}_{\ell m}=&4\pi\left(\frac{4\pi}{3}\right)^2\sqrt{\frac{3}{2\pi}}\frac{\sqrt{6}\sigma_{\mathrm{T}}}{10}\sum_{\lambda =-2}^{2}(-1)^\lambda \int_{0}^{\chi_{i}}d\chi\;e^{-\tau(\chi)}\,n_\mathrm{e}(\chi)\,a(\chi)\int \int \frac{d^3\mathbf{k_1}d^3\mathbf{k_2}}{(2\pi)^6} \tilde{u}(\mathbf{k_{1}})\tilde{u}(\mathbf{k_{2}})\nonumber\\
		&\sum_{{L, M}}(-i)^L Y_{L M}(-\hat{\mathbf{k}})\,j_{L}(k\chi)\sum_{p_1,p_2}
		\left(\begin{array}{ccc}
			1& 1 & 2\\ 
			p_1& p_2 & -\lambda
		\end{array}\right)Y_{1p_1}(-\mathbf{{\hat{k}}_1})Y_{1p_2}(-\mathbf{{\hat{k}_2}})\;A^{\lambda L M}_{\ell m}.
	\end{align}
	Now, using the formula:
	\bea
	\label{y_lm_prop}
	Y_{\ell m}(-\mathbf{\hat{k}})=(-1)^{l+m}\,Y^{*}_{\ell -m}(\mathbf{\hat{k}})
	\eea
	and
	\begin{align}
	\left(\begin{array}{ccc}
		\label{m_reln}
		\ell_{1} & \ell_{2} & \ell_3\\ 
		m_1 & m_2 & m_3
	\end{array}\right)\neq 0 \;\;\text{ if $(m_1 +m_2 +m_3=0)\:\:\text{\&}\:\:(|\ell_1-\ell_2|\leq\ell_3\leq \ell_1+\ell_2)$},
	\end{align}
	we get
	\begin{align}
		a^{*}_{\ell m}=&4\pi\left(\frac{4\pi}{3}\right)^2\sqrt{\frac{3}{2\pi}}\frac{\sqrt{6}\sigma_{\mathrm{T}}}{10}\sum_{\lambda =-2}^{2}(-1)^\lambda \int_{0}^{\chi_{i}}d\chi\;e^{-\tau(\chi)}\,n_\mathrm{e}(\chi)\,a(\chi)\int \int \frac{d^3\mathbf{k_1}d^3\mathbf{k_2}}{(2\pi)^6}\, \tilde{u}(\mathbf{k_{1}})\tilde{u}(\mathbf{k_{2}})\nonumber\\
		&\sum_{{L, M}}(i)^L Y^{*}_{L-M}(\hat{\mathbf{k}})\,j_{L}(k\chi)\sum_{p_1,p_2}(-1)^{(M+\lambda)}
		\left(\begin{array}{ccc}
			1& 1 & 2\\ 
			p_1& p_2 & -\lambda
		\end{array}\right)Y^{*}_{1-p_1}(\mathbf{{\hat{k}}_1})Y^{*}_{1-p_2}(\mathbf{{\hat{k}_2}})\;A^{\lambda L M}_{\ell m}.
	\end{align}
	Again, since $M$, $\lambda $, $p_1$, and $p_2$ are dummy variables which run from $-L\rightarrow L$, $-2\rightarrow2$, $-1\rightarrow1$, and  $-1\rightarrow1$ respectively, we can change $M$ to $-M$, $\lambda$ to $-\lambda$, $p_1$ to $-p_1$, and finally $p_2$ to $-p_2$ without changing the final results. Finally we get
	\begin{align}
		a^{*}_{\ell m}=&4\pi\left(\frac{4\pi}{3}\right)^2\sqrt{\frac{3}{2\pi}}\frac{\sqrt{6}\sigma_{\mathrm{T}}}{10}\sum_{\lambda =-2}^{2}(-1)^\lambda \int_{0}^{\chi_{i}}d\chi\;e^{-\tau(\chi)}\,n_\mathrm{e}(\chi)\,a(\chi)\int \int \frac{d^3\mathbf{k_1}d^3\mathbf{k_2}}{(2\pi)^6} \, \tilde{u}(\mathbf{k_{1}})\tilde{u}(\mathbf{k_{2}})\nonumber\\
		&\sum_{{L, M}}(i)^L Y^{*}_{LM}(\hat{\mathbf{k}})\,j_{L}(k\chi)\sum_{p_1,p_2}(-1)^{(M+\lambda)}
		\left(\begin{array}{ccc}
			1& 1 & 2\\ 
			p_1& p_2 & -\lambda
		\end{array}\right)Y^{*}_{1p_1}(\mathbf{{\hat{k}}_1})Y^{*}_{1p_2}(\mathbf{{\hat{k}_2}})\;A^{-\lambda L -M}_{\ell m}.
	\end{align}
	Next, we use the following properties of the Wigner 3j symbols to simplify further,
	\bea
\left(\begin{array}{ccc}
	l_1& l_2& l_3\\ 
	-m_1& -m_2 & m_3
\end{array}\right)=(-1)^{(l_1+l_2+l_3)}	
\left(\begin{array}{ccc}
	l_1& l_2& l_3\\ 
	m_1& m_2 & -m_3
\end{array}\right).
\eea
		
		Note, that there are Winger 3j symbols in $A^{\lambda L M}_{\ell m}$, which needs to be simplified too. After simplification we get,
		\begin{align}
			a^{*}_{\ell m}=&4\pi\left(\frac{4\pi}{3}\right)^2\sqrt{\frac{3}{2\pi}}\frac{\sqrt{6}\sigma_{\mathrm{T}}}{10}\sum_{\lambda =-2}^{2}(-1)^\lambda \int_{0}^{\chi_{i}}d\chi\;e^{-\tau(\chi)}\,n_\mathrm{e}(\chi)\,a(\chi)\int \int \frac{d^3\mathbf{k_1}d^3\mathbf{k_2}}{(2\pi)^6}\, \tilde{u}(\mathbf{k_{1}})\tilde{u}(\mathbf{k_{2}})\nonumber\\
			&\sum_{{L, M}}(i)^L Y^{*}_{LM}(\hat{\mathbf{k}})\,j_{L}(k\chi)\sum_{p_1,p_2}(-1)^{(L+\ell+m)}
			\left(\begin{array}{ccc}
				1& 1 & 2\\ 
				p_1& p_2 & -\lambda
			\end{array}\right)Y^{*}_{1p_1}(\mathbf{{\hat{k}}_1})Y^{*}_{1p_2}(\mathbf{{\hat{k}_2}})\;A^{\lambda L M}_{\ell -m}.
		\end{align}
		Now, to get the E and B mode coefficients we use  eq.(\ref{e_b_coeff}). Therefore, for E mode we get
		\begin{align}
			\hspace{-0.2cm}e_{\ell m}=&\frac{1}{2}(4\pi)\left(\frac{4\pi}{3}\right)^2\sqrt{\frac{3}{2\pi}}\frac{\sqrt{6}\sigma_{\mathrm{T}}}{10}\sum_{\lambda =-2}^{2}(-1)^\lambda \int_{0}^{\chi_{i}}d\chi\;e^{-\tau(\chi)}\,n_\mathrm{e}(\chi)\,a(\chi) \int \int \frac{d^3\mathbf{k_1}d^3\mathbf{k_2}}{(2\pi)^6} \tilde{u}(\mathbf{k_{1}})\tilde{u}(\mathbf{k_{2}})\nonumber\\
			&\sum_{{L, M}}i^L Y_{L M}^{*}(\hat{\mathbf{k}})\,j_{L}(k\chi)\sum_{p_1,p_2}
			\left(\begin{array}{ccc}
				1& 1 & 2\\ 
				p_1& p_2 & -\lambda
			\end{array}\right)Y_{1p_1}^{*}(\mathbf{{\hat{k}}_1})Y_{1p_2}^{*}(\mathbf{{\hat{k}_2}})\;A^{\lambda L M}_{\ell m}\left(1+(-1)^{(L+\ell)}\right)
		\end{align}
		and similarly for B mode,
		\begin{align}
			\hspace{-0.2cm}b_{\ell m}=&-\frac{i}{2}(4\pi)\left(\frac{4\pi}{3}\right)^2\sqrt{\frac{3}{2\pi}}\frac{\sqrt{6}\sigma_{\mathrm{T}}}{10}\sum_{\lambda =-2}^{2}(-1)^\lambda \int_{0}^{\chi_{i}}d\chi\;e^{-\tau(\chi)}\,n_\mathrm{e}(\chi)\,a(\chi)\int \int \frac{d^3\mathbf{k_1}d^3\mathbf{k_2}}{(2\pi)^6}\tilde{u}(\mathbf{k_{1}})\tilde{u}(\mathbf{k_{2}}) \nonumber\\
			&\sum_{{L, M}}i^L Y_{L M}^{*}(\hat{\mathbf{k}})\,j_{L}(k\chi)\sum_{p_1,p_2}
			\left(\begin{array}{ccc}
				1& 1 & 2\\ 
				p_1& p_2 & -\lambda
			\end{array}\right)Y_{1p_1}^{*}(\mathbf{{\hat{k}}_1})Y_{1p_2}^{*}(\mathbf{{\hat{k}_2}})\;A^{\lambda L M}_{\ell m}\left(1-(-1)^{(L+\ell)}\right).
		\end{align}
		\section{Calculation of the 4-point function \label{App:4point}}
		In this section we will show that out of the three terms that we get when we break the 4 point function using Isserlis theorem, only one term contributes. We begin with eq.(\ref{corr_expansion_0}).
		\begin{align}
			\Big\langle \tilde{u}(\mathbf{k_{1}})\tilde{u}(\mathbf{k_{2}})\tilde{u}^{*}(\mathbf{k_{1}'})\tilde{u}^{*}(\mathbf{k_{2}'})\Big\rangle=&\Big\langle \tilde{u}(\mathbf{k_{1}})\tilde{u}(\mathbf{k_{2}})\Big\rangle\Big\langle \tilde{u}^{*}(\mathbf{k_{1}'})\tilde{u}^{*}(\mathbf{k_{2}'})\Big\rangle+\Big\langle \tilde{u}(\mathbf{k_{1}}) u^{*}(\mathbf{k_{1}'})\Big\rangle\Big\langle \tilde{u}(\mathbf{k_{2}})u^{*}(\mathbf{k_{2}'})\Big\rangle\nonumber\\
			&+\Big\langle \tilde{u}(\mathbf{k_{1}}) u^{*}(\mathbf{k_{2}'})\Big\rangle\Big\langle \tilde{u}(\mathbf{k_{2}})u^{*}(\mathbf{k_{1}'})\Big\rangle.
		\end{align}
		Let us look at the first term. 
		\begin{align}
			\label{corr_1st_term}
			\Big\langle u(\mathbf{k_{1}})u(\mathbf{k_{2}})\Big\rangle\Big\langle u^{*}(\mathbf{k_{1}'})u^{*}(\mathbf{k_{2}'})\Big\rangle&=(2\pi)^6\;P_{uu}(k_1)P_{uu}(k_1')\;\delta(\mathbf{k_1}+\mathbf{k_2})\;\delta(\mathbf{k_1'}+\mathbf{k_2'}).
		\end{align}
		Therefore, using  eq.(\ref{corr_1st_term}) in the angular integral part over $k_1$, $k_2$, $k_1'$, and $k_2'$ of eq.(\ref{cl_ee_1}) and eq.(\ref{cl_bb_1}) we get
		\begin{align}
			&\int d\Omega_{\mathbf{k_{1}}}d\Omega_{\mathbf{k_{1}'}}\;Y^{*}_{1p_1}(\mathbf{{\hat{k}}_1})Y^{*}_{1p_2}(-\mathbf{{\hat{k}_1}})\;Y_{1p_1'}(\mathbf{{\hat{k}}_1'})Y_{1p_2'}(-\mathbf{{\hat{k}_1'}}),\nonumber\\
			&\hspace{1.6in}=(-1)^{(1+p_2)}\int d\Omega_{\mathbf{k_{1}}}\;Y^{*}_{1p_1}(\mathbf{{\hat{k}}_1})Y_{1-p_2}(\mathbf{{\hat{k}_1}})\times\nonumber\\
			&\hspace{1.8in}(-1)^{(1+p_2')}\int d\Omega_{\mathbf{k_{1}'}}Y_{1p_1'}(\mathbf{{\hat{k}}_1'})Y^{*}_{1-p_2'}(\mathbf{{\hat{k}_1'}}),\\
			&\hspace{1.6in}=(-1)^{(p_2+p_2')}\:\delta_{p_1,-p_2}\;\delta_{p_1',-p_2'}.
		\end{align}
		But we see from the Wigner 3j coefficients present in eq.(\ref{cl_ee_1}) and eq.(\ref{cl_bb_1}) that,
		\begin{align}
			\sum_{p_1,p_2}C^{2 \lambda }_{1p_1 1p_2}\;\delta_{p_1,-p_2}
			=\sum_{p_1}C^{2 \lambda }_{1p_1 1-p_1}=0.
		\end{align}
		Hence, the contribution from the first term is zero.
		\section{Quadratic dependence of  the polarisation field on electron's transverse velocity \label{App:Quad_dep}}
		We begin with eq.(\ref{ksz_temp}). As we have shown using Taylor expansion, 
		\bea
		\theta(\mathbf{  \hat{n}'})=\frac{1}{2}v^2-\mathbf{  v}\cdot\mathbf{  \hat{n}'}+\left(\mathbf{  v}\cdot\mathbf{  \hat{n}'}\right)^2+\mathcal{O}\left(\left(\mathbf{  v}\cdot\mathbf{  \hat{n}'}\right)^3\right)+\cdot\cdot\cdot
		\eea
		From eq.(\ref{k_quadrupole}), we saw that the contribution to the intensity for the SZ part of the spectrum is proportional to $\left(\theta(\mathbf{  \hat{n}'})\right)^2$. Therefore squaring and rearranging the terms we get:
		\bea
		\left(\theta(\mathbf{  \hat{n}'})\right)^2=\left(\mathbf{  v}\cdot\mathbf{  \hat{n}'}\right)^2-\frac{1}{3}v^2+ \frac{1}{3}v^2 +\mathcal{O}(v^4)
		+\cdots
		\eea
		Therefore, the contribution to the quadrupolar moment will just be  $\frac{1}{3}v^2\left(3\cos^2\zeta-1\right)$, where $\zeta$ is the angle between $\mathbf{v}$ and $\mathbf{\hat{n}'}$. The monopole term $\propto v^2$ will not contribute anyway when we integrate over $d^{2}\mathbf{ \hat{n}'}$. Now, we can write,
		\bea
		\frac{1}{3}v^2\left(3\cos^2\zeta-1\right)=\frac{4}{3}\sqrt{\frac{\pi}{5}}\;v^2\;Y_{20}(\hat{\mathbf{v}};\mathbf{\hat{n}'}),
		\eea
		where $Y_{20}(\hat{\mathbf{v}};\mathbf{\hat{n}'})$ is defined by considering the z-direction to be along $\mathbf{\hat{n}'}$. Now we can use the following property of spherical harmonics to split it as a product of two spherical harmonics \cite{durrer2020cosmic},
		\bea
		Y_{20}(\hat{\mathbf{v}};\mathbf{\hat{n}'})=\sqrt{\frac{4\pi}{5}}\sum_{m'}\;Y^{*}_{2m'}(\hat{\mathbf{v}};\mathbf{\hat{e}})\,Y_{2m'}(\mathbf{\hat{n}'};\mathbf{\hat{e}}),
		\eea
		where $\mathbf{\hat{e}}$ is some general direction which is our new z-direction. Therefore, from eq.(\ref{pol_sem_final}) we observe
		\bea
		\int d^{2}\mathbf{ \hat{n}'}\;Y_{2 \lambda}^{*}\left(\mathbf{ \hat{n}'}\right)\mathcal{I}_{\mathrm{sc}}\left(\mathbf{r},\mathbf{  \hat{n}'}\right)= \frac{8\pi}{15}\;v^2\;Y^{*}_{2\lambda}(\hat{\mathbf{v}};\mathbf{\hat{e}}).
		\eea
		So, finally we get,
		\begin{align}	
		P_{+}\left(\hat{\mathbf{ n}}\right)&=- \frac{\sqrt{6}\sigma_{\mathrm{T}}}{10}\sum_{\lambda =-2}^{2}\int_{0}^{\chi_{i}}d\chi\;e^{-\tau(\chi)}\,n_\mathrm{e}(\chi)a(\chi)\,_{2}Y_{2 \lambda}\left(\mathbf{ \hat{n}}\right)\left[\frac{8\pi}{15}\;v^2\;Y^{*}_{2\lambda}(\hat{\mathbf{v}};\mathbf{\hat{e}})\right],\\
			&=-\frac{8\pi}{15}\frac{\sqrt{6}\sigma_{\mathrm{T}}}{10}\int_{0}^{\chi_{i}}d\chi\;e^{-\tau(\chi)}\,n_\mathrm{e}(\chi)a(\chi)v^2\left[\sum_{\lambda =-2}^{2}\,_{2}Y_{2 \lambda}\left(\mathbf{ \hat{n}}\right)\;Y^{*}_{2\lambda}(\hat{\mathbf{v}};\mathbf{\hat{e}})\right],\\
			&=-\frac{8\pi}{15}\frac{\sqrt{6}\sigma_{\mathrm{T}}}{10}\int_{0}^{\chi_{i}}d\chi\;e^{-\tau(\chi)}\,n_\mathrm{e}(\chi)a(\chi)v^2\sqrt{\frac{5}{4\pi}}\;Y_{2-2}(\mathbf{\hat{v}};\mathbf{\hat{n}}),
		\end{align}
To simplify this further we consider $\hat{\mathbf{n}}$ as the new $\hat{\mathbf{z}}$ direction. Doing so, we can split the spherical harmonics $Y_{2-2}(\mathbf{\hat{v}};\mathbf{\hat{n}})$ in terms of angle $\theta$ and $\phi$ of the spherical polar coordinate system. Thus we get
	\begin{align}
		P_{+}{\left(\hat{\mathbf{n}}\equiv \hat{\mathbf{z}}\right)}&=-\frac{\sigma_{\mathrm{T}}}{10}\int_{0}^{\chi_{i}}d\chi\;e^{-\tau(\chi)}\,n_\mathrm{e}(\chi)a(\chi)v^2\,\sin^2\theta\;e^{-2i\phi},\\
			&=-\frac{\sigma_{\mathrm{T}}}{10}\int_{0}^{\chi_{i}}d\chi\;e^{-\tau(\chi)}\,n_\mathrm{e}(\chi)a(\chi)v_t^2\;e^{-2i\phi},
		\end{align}
		where $v_t=v\sin\theta$ is the transverse to $\hat{\mathbf{ n}}$.
		Thus, we have shown that the polarisation signal is proportional to the square of the transverse velocity field.
		\section{Electron number density profile for galaxy clusters \label{App:E_density}}
		We considered a Gaussian profile for the gas present in the ICM. One may consider a more realistic profile, such as given in \cite{2002MNRAS.336.1256K}, but for scales much larger than the halo sizes, the exact nature of the profile is unimportant. What matters is the volume occupied by the gas regardless of the detailed shape. This makes the calculation faster without decreasing the accuracy of our final results. We start with a halo profile of the form
		\begin{align}
			\rho(r)=\rho_0\exp\left(-\frac{4r^2}{R^2}\right).
		\end{align}
		where R is some scale radius. For a halo with mass = $\mathcal{M}$, the scale radius is so chosen that $R(\mathcal{M}_\mathrm{200m}) =0.95\mathcal{M}$. Using the normalisation, $\int 4\pi\rho(r)r^2dr=\mathcal{M}$, we get $\rho_0$ as
		\bea
		\rho_0=\frac{8\mathcal{M}}{\pi^{3/2}R^3}.
		\eea  
		The gas density can be written as 
		\bea 
		\rho_{\mathrm{gas}}(r)=\frac{\Omega_b}{\Omega_m}\rho(r)=\frac{\Omega_b}{\Omega_m}\frac{8\mathcal{M}}{\pi^{3/2}R^3}\exp\left(\frac{-4r^2}{R^2}\right).
		\eea
		From the gas density we can easily find the electron number density by dividing it by mean gas mass per electron 
		\begin{align}
			n_\mathrm{e}(r)=\frac{\rho_{\mathrm{gas}}(r)}{1.14m_\mathrm{p}}&=\frac{\Omega_b}{\Omega_m}\frac{1}{1.14\pi^{3/2}}\frac{8\mathcal{M}}{m_\mathrm{p}R^3}\exp\left(\frac{-4r^2}{R^2}\right),\\
			&=n_\mathrm{e}^{0}\;W(r).
		\end{align}  
		where $n_\mathrm{e}^{0}=\frac{\Omega_b}{\Omega_m}\frac{1}{1.14\pi^{3/2}}\frac{8\mathcal{M}}{m_\mathrm{p}R^3}$, $W(r)=\exp\left(\frac{-4r^2}{R^2}\right)$ and $m_\mathrm{p}$ is the mass of proton. So if we take the Fourier transform of the electron number density we get
		\begin{align}
			n_\mathrm{e}(k)&=n_\mathrm{e}^{0}\int d\mathbf{r}\exp\left(-i\mathbf{k}\cdot\mathbf{r}\right)W(r),\nonumber\\
			&=n_\mathrm{e}^{0} \int dr\,4\pi r^2\;\frac{\sin(kr)}{kr}\exp\left(\frac{-4r^2}{R^2}\right),\nonumber\\
			&=n_\mathrm{e}^{0}\frac{R^3}{8}\,\pi^{3/2}\exp\left(\frac{-k^2\,R^2}{16}\right),\nonumber\\
			&=\frac{\Omega_b}{1.14\Omega_m}\frac{\mathcal{M}}{m_\mathrm{p}}\exp\left(\frac{-k^2\,R^2}{16}\right).
		\end{align} 	
	\section{Poisson contribution to polarisation power spectra. \label{App:Poisson}}
	In this section, we derive the power spectrum of E and B modes from the Poisson term.  We have to repeat the same process as shown in the case of uniform electron number density field to obtain the power spectra. The density-density correlations and the velocity-velocity correlations can be calculated separately. Using eq.(\ref{e_no_den_cluster}) in place of electron number density in eq.(\ref{pol_final}) and doing a variable change from conformal time to comoving distance as shown earlier, we get in temperature units
			\begin{align}
				\hspace{0cm}
			\label{cl_ee_poi}
			C^{EE\,(\mathrm{Poi})}_{\ell}=&	\frac{T^{2}_{\mathrm{CMB}}}{2}\left[(4\pi)\left(\frac{4\pi}{3}\right)^2\sqrt{\frac{3}{2\pi}}\frac{\sqrt{6}\sigma_{\mathrm{T}}}{10}n_\mathrm{e}^{0}\right]^2\sum_{\lambda,\lambda' =-2}^{2}(-1)^{(\lambda+\lambda')}\int_{0}^{\chi_{i}}d\chi\;e^{-\tau(\chi)}\, a(\chi)\nonumber\\
			&\int_{0}^{\chi_{i}}d\chi'e^{-\tau(\chi')}\,a(\chi')\int d\mathcal{M}\,\bar{n}(\mathcal{M},\chi)\sum_{{L, M}\atop{L',M'}}\sum_{{p_1,p_2}\atop{p_1',p_2'}}i^{(L-L')}
			\left(\begin{array}{ccc}
				1& 1 & 2\\ 
				p_1& p_2 & -\lambda
			\end{array}\right)
		\left(\begin{array}{ccc}
			1& 1 & 2\\ 
			p_1'& p_2' & -\lambda'
		\end{array}\right)\nonumber\\
			&\int \int \frac{dk_1dk_2}{(2\pi)^6}k^2_{1}k^2_{2}\; P_{uu}(k_1)P_{uu}(k_2)\int \frac{dk_3k^2_{3}}{(2\pi)^3}\mathrm{W}\left(k_3,\mathcal{M}\right)^2j_{L}(k_3\chi)\,j_{L'}(k_3\chi')\nonumber\\
			&\int d\Omega_{\mathbf{k_1}}\;Y_{1p_1}^{*}(\mathbf{{\hat{k}}_1})Y_{1p_1'}(\mathbf{{\hat{k}}_1})\int d\Omega_{\mathbf{k_2}}Y_{1p_2}^{*}(\mathbf{{\hat{k}_2}})Y_{1p_2'}(\mathbf{{\hat{k}_2}})\int d\Omega_{\mathbf{k_3}}\;Y_{L M}^{*}(\mathbf{\hat{k}_3}) Y_{L' M'}(\mathbf{\hat{k}_3})\,\times\nonumber\\
			&\hspace{1.5in}A^{\lambda L M}_{\ell m}A^{\lambda' L' M'}_{\ell m}\left(1+(-1)^{(L+\ell)}\right)\left(1+(-1)^{(L'+\ell)}\right)
		\end{align}
		and similarly,
		\begin{align}
			\label{cl_bb_poi}
			C^{BB\,(\mathrm{Poi})}_{\ell}=&	\frac{T^{2}_{\mathrm{CMB}}}{2}\left[(4\pi)\left(\frac{4\pi}{3}\right)^2\sqrt{\frac{3}{2\pi}}\frac{\sqrt{6}\sigma_{\mathrm{T}}}{10}\,n_\mathrm{e}^{0}\right]^2\sum_{\lambda,\lambda' =-2}^{2}(-1)^{(\lambda+\lambda')}\int_{0}^{\chi_{i}}d\chi\;e^{-\tau(\chi)}\, a(\chi)\nonumber\\
			&	\int_{0}^{\chi_{i}}d\chi'e^{-\tau(\chi')}\,a(\chi')\int d\mathcal{M}\,\bar{n}(\mathcal{M},\chi)\sum_{{L, M}\atop{L',M'}}\sum_{{p_1,p_2}\atop{p_1',p_2'}}i^{(L-L')}	
			\left(\begin{array}{ccc}
				1& 1 & 2\\ 
				p_1& p_2 & -\lambda
			\end{array}\right)
			\left(\begin{array}{ccc}
				1& 1 & 2\\ 
				p_1'& p_2' & -\lambda'
			\end{array}\right)\nonumber\\
			&\int \int \frac{dk_1dk_2}{(2\pi)^6}k^2_{1}k^2_{2}\;P_{uu}(k_1)P_{uu}(k_2) \int \frac{dk_3k^2_{3}}{(2\pi)^3}\mathrm{W}\left(k_3,\mathcal{M}\right)^2j_{L}(k_3\chi)\,j_{L'}(k_3\chi')\nonumber\\
			&\int d\Omega_{\mathbf{k_1}}\;Y_{1p_1}^{*}(\mathbf{{\hat{k}}_1})Y_{1p_1'}(\mathbf{{\hat{k}}_1}) \int d\Omega_{\mathbf{k_2}}Y_{1p_2}^{*}(\mathbf{{\hat{k}_2}})Y_{1p_2'}(\mathbf{{\hat{k}_2}})\int d\Omega_{\mathbf{k_3}}\;Y_{L M}^{*}(\mathbf{\hat{k}_3}) Y_{L' M'}(\mathbf{\hat{k}_3})\,\times\nonumber\\
			&\hspace{1.5in}A^{\lambda L M}_{\ell m}A^{\lambda' L' M'}_{\ell m}\left(1-(-1)^{(L+\ell)}\right)\left(1-(-1)^{(L'+\ell)}\right).
		\end{align}
		The angular integrals can be performed analytically in this case. We can also sum over the Wigner 3j symbols. After doing these simplifications, we finally get,
		\begin{align}
			\label{cl_ee_poi_final}
			C^{EE\,(\mathrm{Poi})}_{\ell}=&T^{2}_{\mathrm{CMB}}\,	(2\pi)\left[\left(\frac{4\pi}{3}\right)^2\sqrt{\frac{3}{2\pi}}\frac{\sqrt{6}\sigma_{\mathrm{T}}}{10}\,n_\mathrm{e}^{0}\right]^2\int_{0}^{\chi_{i}}d\chi\;e^{-\tau(\chi)}\, a(\chi)\int_{0}^{\chi_{i}}d\chi'\;e^{-\tau(\chi')}\, a(\chi')\nonumber\\
			&\hspace{0.1in}\int d\mathcal{M}\,\bar{n}(\mathcal{M},\chi)\sum_{L}(2L+1)
			\left[	\left(\begin{array}{ccc}
				L& 2& \ell\\ 
			   0& -2 & 2
			\end{array}\right)\right]^2\int \int \frac{dk_1dk_2}{(2\pi)^6}\;k^2_{1}k^2_{2}P_{uu}(k_1) P_{uu}(k_2)\nonumber\\
			&\hspace{0.9in}\int \frac{dk_3k^2_{3}}{(2\pi)^3}\mathrm{W}\left(k_3,\mathcal{M}\right)^2j_{L}(k_3\chi)\,j_{L}(k_3\chi')\;\left(1+(-1)^{(L+\ell)}\right)^2
		\end{align}
		and similarly,
		\begin{align}
			\label{cl_bb_poi_final}
			C^{BB\,(\mathrm{Poi})}_{\ell}=&T^{2}_{\mathrm{CMB}}\,	(2\pi)\left[\left(\frac{4\pi}{3}\right)^2\sqrt{\frac{3}{2\pi}}\frac{\sqrt{6}\sigma_{\mathrm{T}}}{10}\,n_\mathrm{e}^{0}\right]^2\int_{0}^{\chi_{i}}d\chi\;e^{-\tau(\chi)}\, a(\chi)\int_{0}^{\chi_{i}}d\chi'\;e^{-\tau(\chi')}\, a(\chi')\nonumber\\
			&\hspace{0.1in}\int d\mathcal{M}\,\bar{n}(\mathcal{M},\chi)\sum_{L}(2L+1)
				\left[	\left(\begin{array}{ccc}
				L& 2& \ell\\ 
				0& -2 & 2
			\end{array}\right)\right]^2\int \int \frac{dk_1dk_2}{(2\pi)^6}\;k^2_{1}k^2_{2}P_{uu}(k_1)P_{uu}(k_2)\nonumber\\
			&\hspace{0.9in} \int \frac{dk_3k^2_{3}}{(2\pi)^3}\mathrm{W}\left(k_3,\mathcal{M}\right)^2j_{L}(k_3\chi)\,j_{L}(k_3\chi')\;\left(1-(-1)^{(L+\ell)}\right)^2.
		\end{align}
	We choose the mass range of the clusters to be between $10^{13} \mathrm{M_{\odot}}$ to $10^{17} \mathrm{M_{\odot}}$.

		\bibliographystyle{unsrtads}
		\bibliography{references.bib}

			

	\end{document}